\documentclass[10pt,twocolumn,twoside]{IEEEtran}

\usepackage{graphics}
\usepackage{epstopdf}

\usepackage{amssymb,amsmath,graphicx}
\usepackage{pbox}
\usepackage{ragged2e}
\usepackage{url}
\usepackage[table]{xcolor}

\usepackage[table]{xcolor}
\usepackage{subcaption,multirow}
\usepackage{float}
\usepackage[table]{xcolor} 
\usepackage{soul}
\usepackage{hyperref}
\usepackage{arydshln}
\usepackage{lipsum}

\usepackage{fancyhdr}

\begin{document}

% \title{Time-Contrastive Learning and Phone-Discrimination Based Bottleneck Features for Text-Dependent Speaker Verification}
\title{Time-Contrastive Learning Based Deep Bottleneck Features for Text-Dependent Speaker Verification}

\author{Achintya kr. Sarkar, Zheng-Hua Tan, \emph{Senior Member, IEEE}, Hao Tang, Suwon Shon and \\ James Glass, \emph{Fellow, IEEE} % <-this % stops a space
\thanks{This work of A. K. Sarkar and Z.-H. Tan was supported by the iSocioBot project, funded by the Danish Council for Independent
 Research - Technology and Production Sciences (\#1335-00162). (\emph{Corresponding author: Z.-H. Tan}) 

A. K. Sarkar is with the School of Electronics Engineering, VIT-AP University, India. This work was done while A. K. Sarkar was post-doctoral research fellow at the Department of Electronic systems, Aalborg University, 9220 Aalborg, Denmark.

Z.-H. Tan is with the Department of Electronic systems, Aalborg University, 9220 Aalborg, Denmark. This work was done in part while Z.-H. Tan was visiting Computer Science and Artificial Intelligence Laboratory, Massachusetts Institute of Technology, Cambridge MA, USA. 

H. Tang, S. Shon and J. Glass are with Computer Science and Artificial Intelligence Laboratory, Massachusetts Institute of Technology, Cambridge, MA 02139, USA.
}
}

\makeatletter
%%%%%%%%%%%%%%%%%%%%%%%%%%%%%% User specified LaTeX commands.
\def\ps@IEEEtitlepagestyle{%
  \def\@oddfoot{\mycopyrightnotice}%
  \def\@evenfoot{}%
}
\def\mycopyrightnotice{%
  {\footnotesize Copyright $\copyright$ 2019 IEEE. This article is the \emph{accepted} version of IEEE/ACM-TASLP.
DOI:10.1109/TASLP.2019.2915322. \hfill}% <--- Change here
  \gdef\mycopyrightnotice{}% just in case
}

\hypersetup{plainpages = false,
              breaklinks=true,
              %a4paper=true,
              linktocpage,
              colorlinks   = true,  %Colours links instead of ugly boxes
              urlcolor     = blue,  %Colour for external hyperlinks
              linkcolor    = blue,  %Colour of internal links
              citecolor   = blue,    %Colour of citations
              pdfstartpage=1,
              pdfstartview=FitH,
              pdfkeywords={}}

%\begin{document}

\maketitle

\begin{abstract}
There are a number of studies about extraction of bottleneck (BN) features from deep neural networks (DNNs) trained to discriminate speakers, pass-phrases and triphone states for improving the performance of text-dependent speaker verification (TD-SV). However, a moderate success has been achieved. A recent study \cite{Aapo2016} presented a time contrastive learning (TCL) concept to explore the non-stationarity of brain signals for classification of brain states. Speech signals have similar non-stationarity property, and TCL further has the advantage of having no need for labeled data. We therefore present a TCL based BN feature extraction method. The method uniformly partitions each speech utterance in a training dataset into a predefined number of multi-frame segments. Each segment in an utterance corresponds to one class, and class labels are shared across utterances. DNNs are then trained to discriminate all speech frames among the classes to exploit the temporal structure of speech. 
In addition, we propose a segment-based unsupervised clustering algorithm to re-assign class labels to the segments. TD-SV experiments were conducted on the RedDots challenge database. The TCL-DNNs were trained using speech data of fixed pass-phrases that were excluded from the TD-SV evaluation set, so the learned features can be considered phrase-independent. We compare the performance of the proposed TCL bottleneck (BN) feature with those of short-time cepstral features and BN features extracted from DNNs discriminating speakers, pass-phrases, speaker+pass-phrase, as well as monophones whose labels and boundaries are generated by three different automatic speech recognition (ASR) systems.  Experimental results show that the proposed TCL-BN outperforms cepstral features and speaker+pass-phrase discriminant BN features, and its performance is on par with those of ASR derived BN features. Moreover, the clustering method improves the TD-SV performance of TCL-BN and ASR derived BN features with respect to their standalone counterparts. We further study the TD-SV performance of fusing cepstral and BN features. 

% \textcolor{red}{, i.e. to learn features discriminating the temporal non-stationary structure of speech in unsupervised manner.}
% As speakers in the TD-SV are constrained to speak the same text/pass-phrase during the training and test phase. Hence, TD-SV maintains  the  phonetically/temporal  matched contents between the training and test phases. So, the TCL-DNNs were trained using speech data of fixed pass-phrases that were excluded from the TD-SV evaluation set, so the learned features can be considered phrase-independent. %and is superior or similar to those of ASR derived BN features, depending on their corresponding ASR performance.  
% augmenting cepstral features with BN features, and improvements are observed. 
% It demonstrates the effectiveness of the proposed TCL-BN in TD-SV with respect to the existing BN features. 

\end{abstract}

\section{Introduction}
Due to the quasi-periodic nature of speech, short-time acoustic cepstral features are widely used in speech and speaker recognition. Recent development of deep neural networks (DNNs) \cite{Hinton2012} has ignited a great interest in using bottleneck (BN) features \cite{Fu2014,yu2017adversarial,Yuan2015,Ghalehjegh2015,Cong-Thanh2013, Variani2014,Yaman2012} for speech classification tasks including speaker verification (SV). 
The goal of SV is to verify a person using their voice \cite{reynold00, kinnunen2010overview}. SV methods can be broadly divided into text-dependent (TD) and text-independent (TI) ones \cite{sarkar2018incorporating}. 
In TD-SV, speakers are constrained to speak the same pass-phrase or sentence during both enrolment and test phases. 
In TI-SV, speakers can speak any sentence during enrolment and test phases, i.e. there is no constraint on what sentences to be uttered. Since TD-SV makes use of a matched phonetic content during enrolment and test phases, it typically outperforms TI-SV.

A classical speaker verification system in general involves discriminative feature extraction, universal background modelling, and training of Gaussian mixture model-universal background model (GMM-UBM) or i-vector, which is a fixed- and low-dimensional representation of a speech utterance \cite{Deka_ieee2011}. DNNs are applied to SV in all these three parts: 1) extracting discriminative bottleneck features \cite{Yuan2015}, 2) replacing GMM-UBM for i-vector extraction \cite{mclaren2015advances}, and 3) directly replacing i-vectors with speaker embeddings \cite{Li2018}, in addition to works aiming to improve SV robustness against noise \cite{michelsanti2017conditional, eskimez2018front} and domain variation \cite{wang2018unsupervised}. When used for replacing UBM, a DNN that is trained as an acoustic model of automatic speech recognition (ASR) replaces the traditional GMM-UBM by predicting posteriors of senones (e.g., tied-triphone states). This allows to incorporate phonetic knowledge into i-vectors. DNNs are also used to directly replace i-vectors for speaker characterization with trained speaker embeddings, which are the outputs of one or more DNN hidden layers. In \cite{variani2014deep}, the embeddings are also called d-vector. Instead of equally weighting and averaging all frames as e.g. in the d-vector approach, paper \cite{zhang2016end} uses an attention mechanism to fuse phonetic and speaker representations so as to generate an utterance-level speaker representation. When used for feature extraction, a DNN is trained to discriminate speakers, pass-phrases, senones or a combination of them.
Then the outputs of one or more DNN hidden layers are projected onto a low dimensional space called BN features.
Previous studies \cite{Yuan2015,Ghalehjegh2015,Cong-Thanh2013,DBLP:journals/spl/SarkarDLB14,Shi2018, Ranjan2017} have demonstrated that BN features are useful either for obtaining a better performance than cepstral features or for providing complementary information when cepstral and BN features are fused. 
% The study in \cite{Yuan2015} has shown that BN features based on discriminating speakers or speaker+phone or speaker+pass-phrase are able to reduce TD-SV error rates when compared to conventional short-time cepstral feature. 
However, training DNNs to extract these BN features requires manual labels (e.g., speakers and pass-phrases), or phonetic transcriptions based on ASR. Obtaining these labels are time-consuming and expensive, and building ASR systems requires large amounts of training data and expert knowledge \cite{harwath2016unsupervised}. Beyond SV, some other works extract phonetic annotation based BN features for speech recognition \cite{DBLP:conf/asru/ChenLXML17,DBLP:conf/asru/CYuan17} and spoken language recognition \cite{DBLP:conf/asru/ChenLXML17, DBLP:conf/interspeech/LiMBPD16, Radek17}.

Unsupervised representation learning is one of the biggest challenges in machine learning and at the same time has a great potential of leveraging the vast amount of often unlabeled data. The primary approach to unsupervised deep learning is probabilistic generative modeling, due to optimal learning objectives that probabilistic theory is able to provide \cite{Aapo2016}, \cite{hsu2017unsupervised}. Successful examples are variational autoencoders (VAEs) \cite{kingma2013auto} and generative adversarial networks (GANs) \cite{goodfellow2014generative}. The study in \cite{Aapo2016} presents a time contrastive learning (TCL) concept, a type of unsupervised feature learning method, which explores the temporal non-stantionarity of time series data. The learned features aim to discriminate data from different time segments. It is shown that what the TCL feature extractor computes is the log-probability density function of the data points in each segment, and thus TCL has a nice probabilistic interpretation \cite{Aapo2016}. The TCL method is used for classifying a small number of different brain states that generally evolve over the time and can be measured by magnetoencephalography (MEG) signals. Specifically, TCL trains a neural network to discriminate each segment by using the segment indices as labels. The output of the last hidden layer is the feature for classifying brain states \cite{Aapo2016}. % by formulating a generative model where independent components have different distributions in different time windows, and we observe nonlinear mixtures of the components
% \cite{Aapo2016}  However, we showed that, surprisingly, it can estimate independent components in a nonlinear mixing model up to certain indeterminacies, assuming that the independent components are nonstationary in a suitable way. The indeterminacies include a linear mixing (which can be resolved by a further linear ICA step), and component-wise nonlinearities,

Exploiting underlying structure of temporal data for unsupervised feature learning has also been studied for video data. In \cite{goroshin2015unsupervised}, features are learned in an unsupervised fashion by assuming that data points being neighbors in the temporal space are likely to be neighbors in the latent space as well. Similarly, the work in \cite{mobahi2009deep} exploits the structure of video data based on two facts: (1) there is a temporal coherence in two successive frames, namely they contain similar contents and represent the same concept class, and (2) there are differences or changes among neighbouring frames due to, e.g., translation and rotation. Therefore, learning features by exploiting this structure will be able to generate representations that are both meaningful and invariant to theses changes \cite{mobahi2009deep}.

Since speech is a non-stationary time series signal, there is a contrast across speech segments. At the same time, neighbouring frames likely represent the same concept class. Furthermore, since in the TD-SV setting, same pass-phrases are uttered by speakers multiple times in the training set, there are certain structures in the data, e.g. matched contents across utterances. Across the entire training dataset, segments assigned with the same classes are of course most likely heterogeneous. In \cite{zhang2016understanding}, however, it is shown that deep neural networks trained by stochastic gradient descent methods can fit well the training image data with random labels and this phenomenon happens even if the true images are replaced by unstructured random noise. Therefore, we hypothesize that training of networks with random labels assigned by the TCL approach will converge and if we choose bottleneck features from the proper hidden layer, a useful feature can be extracted. All these motivate us to propose the TCL method for TD-SV. Speech and MEG signals, however, are quite different in nature, namely speech signals contain much richer information for which the tasks in hand often involve classification of much more classes. Furthermore, the amount of available speech data including labelled data is significantly larger than MEG data, leading to more alternative methods for speech feature learning. Therefore, extensive study is required to explore the potential of TCL for speech signals. 
% since the speakers in a TD-SV task are constrained to speak the same pass-phrases during the training and test phases, there it maintains  a phonetically/temporal  matched contents between the training the test phase

In \cite{AchintyaNips2017} we proposed a TCL based BN feature for TD-SV. The main strategy is to uniformly partition each utterance into a predefined number of  segments, e.g. $N$, regardless of speakers and contents. The first segment in an utterance is labelled as Class 1, the second as Class 2, and so on. Each segment is assumed to contain a single content belonging to a class. The speech frames within the $n^{th}$ segment, $n\in\{1,2,...,N\}$, are assigned to Class $n$. A DNN is then trained to discriminate each speech frame among the different classes. The core idea of TCL learning is to exploit temporally varying characteristics inherent in speech signals. 
It has been shown in \cite{AchintyaNips2017} that without using any label information for DNN training, TCL-BN gives better TD-SV performance than the Mel-frequency cepstral coefficient (MFCC) feature and existing BN features extracted from DNNs trained to discriminate speakers or both speakers and pass-phrases where manual labels are exploited. % The advantage of TCL is its ability to use the vast amount of unlabelled speech data available in the wild.

While no need for labelled data is an advantage, segmentation and labelling in TCL are arbitrary and the labels do not carry any particular meaning. In this work, we therefore propose a segment-based statistical clustering method to iteratively regroup the segments in an unsupervised manner with the goal to maximize likelihood. The clustering method groups together segments with similar phonetic content to form clusters, and each cluster is considered a class. It is expected that the clustering process will lead to improved class labels for the segments, which are then used to train DNNs, leading to improved BN features. % Experiments on the RedDots Challenge 2016 database show that the clustering method is able to improve TD-SV performance of the TCL method.
% in maximum likelihood (ML) sense. As TCL does not use any label or manual/automatic annotation/transcriptions and hence clustering is expected to help further regroup the similar phonetic class/event on the data within same class in unsupervised manner. In other wards, TCL+clustering will outcome the better class label of the data and hence will yield  better label for the training data of DNN in TCL i.e. expected better optimized DNN and hence better TCL-BN features for the TD-SV. This will finally improve the performance of the overall TD-SV. We further study the performance of the TD-SV for augmentation of cepstral feature with BN  in both score and feature domain. It demonstrates the effectiveness of the proposed TCL-BN in TD-SV with respect to the existing BN features. System performance are represented on the Reddots challenge 2016 database for TD-SV using short utterances  

As the TCL method trains DNNs to discriminate phonetic content, one natural question to ask is how it compares with segmentation and labelling obtained by a speech recognizer.
While senones or triphone states have been used as the target classes for training DNNs to extract features, BN feature extraction based on discriminating phones is relatively unexplored in the context of TD-SV. The motivation of investigating the use of phones is that the time granularity or resolution for defining the classes is significantly smaller than that of using triphone states (e.g. 3001 in \cite{Yuan2015}) and much closer to that of TCL learning (e.g. tens in \cite{AchintyaNips2017}). 
% \ztcomment{This makes a difference as I see it. An experimental comparison is needed.}
In \cite{Yuan2015}, triphone states have been used as the frame labels for training DNNs from which BN features are extracted. It is shown that BN features extracted from DNNs discriminating both speakers and phones performs similarly to BN features based on discrimination of either speakers only or both speakers and phrases. In \cite{mclaren2015advances}, bottleneck features are extracted from DNNs trained to predict senone posteriors. Experimental results show that the senone-discriminant BN feature does not even outperform MFCCs, although being complementary to MFCCs. The reason why using senones as training targets does not improve the MFCC baseline might be because the large number of senones requires to use a large amount of data to train a large neural network in order to perform well. Instead of using tied tri-phone states/senones as the DNN training targets as in \cite{Yuan2015,mclaren2015advances}, this paper investigates two speech recognition settings, one where a phoneme recognizer is used to decode the phone sequences, for which two different recognizers are investigated, and the other where the forced alignments are used to obtain the phone sequences. The generated phone sequences and boundaries are used for training phone-discriminant BN (PHN-BN) features. We compare their performance against each other and that of TCL. To our knowledge, the performance of using PHN-BN features for TD-SV has not been reported in the literature. Context-independent monophone states have been used as DNN training targets to extract BN features for language identification in \cite{geng2015multilingual}, where it is experimentally shown that phone-state-discriminant BN performs significantly better than the triphone-state-discriminant BN. However, monophone states rather than phones themselves are used and the application is language identification rather than SV \cite{geng2015multilingual}. 

% However, performance of the TD-SV with TCL-BN  has not compared with the BN system  discriminating phones using ASR generated annotation. Therefore, it is unclear from the study that what is the gap between the TCL-BN system (without annotation) and ASR-BN systems for the TD-SV. In other words, how TCL-BN  competes with the ASR-BN if best possible available automatic machine learning algorithm is used to generate the label for BN to exploit the large amount of data available in the real life. 
%In this paper, we \emph{first} consider different  ASR (e.g. force-alignment) transcriptions based BN features for the comparison of performance of TD-SV with TCL-BN, where BN features are extracted using discriminate data-points among the different phonetic classes.
%  This will demonstrate how extends TCL-BN can meet the performance of the TD-SV with compare to the ASR-BN. %(using the best possible way of getting phonetic transcription for a large amount of DNN training data available in the real-life for machine learning). 
  
We conducted our TD-SV experiments on the RedDots Challenge 2016 database \cite{lee2015reddots, RedDots}. %using short utterances 
We show that TCL-BN gives better performance than MFCC features and BN features discriminating speakers or both speakers and pass-phrases, while being on par with using the phone sequences produced by an ASR system. Clustering improves the performance especially for TCL-BN, and TCL-BN with clustering performs the best among all features. The TCL approach further has the flexibility in choosing the number of target classes for DNN training.

The contributions of this paper are multi-fold. First, it proposes a segment-based statistical clustering method to re-assign class labels to the segments generated by TCL or speech recognizers. Second, the paper extends the study of our previous work on TCL-BN \cite{AchintyaNips2017}, to analyse the learned features through scatter plots using the T-SNE method \cite{tsne} and to conduct more extensive experiments such as extracting BN features from different DNN hidden layers with different numbers of DNN training target classes. Third, the paper studies BN features that are extracted from DNNs trained to discriminate phones, which are again based on segmentation and labeling generated by different ASR systems, in contrast to training DNNs to discriminate triphone states or senones as done in the literature. Fourth, the performance of a wide range of BN features are compared under the GMM-UBM and i-vector frameworks on the RedDots database. Finally, the fusion of MFCCs and various BN features at both score and feature levels is studied. 

The rest of the paper is organized as follows. In Section II we describe bottleneck features. The segment-based clustering method is presented in Section III. Sections IV and V present two TD-SV methods and experimental set-ups, respectively. Results and discussions are given in Section VI. The paper concludes in Section VII. 
% \htcomment{can remove this paragraph if we need space.}
% \ztcomment{agree.}

\section{Bottleneck Features}
Bottleneck features are features extracted from the hidden layers of BN-DNNs (i.e. DNNs for BN feature extraction). In this section, we present three phone-discriminant BN features, which differ from the often used senone-discriminant BN features, and two time-contrastive learning based BN features, in addition to the commonly used speaker- and pass-phase-discriminant BN features.

All BN-DNNs in this work use Mel-frequency cepstral coefficients \cite{Davis80} as the input. MFCCs are the most commonly used features for speaker verification. In this work, we use $57$ dimensional MFCCs including $C_1$-$C_{19}$, $\Delta$ and $\Delta \Delta$ coefficients with RASTA filtering \cite{Hermansky90}, which are extracted from speech signals with a $20$ ms Hamming window and a $10$ ms frame shift. An energy based voice activity detection is applied to select high energy frames for MFCC feature extraction and further processing, while low energy frames are discarded. This work does not consider noisy speech signals and otherwise, it will be essential to use a noise robust voice activity detection method. Finally, the high energy frames are normalized to fit zero mean and unit variance at utterance level.

\subsection{Speaker- and pass-phrase-discriminant BN features}

Two BN features are chosen as state-of-the-art baseline methods in this work. The first one is speaker-discriminant BN (SPK-BN) \cite{Yuan2015}, in which DNNs are trained to discriminate speakers using the cross-entropy loss. 
%, a cross-entropy based objective function is optimized for discriminating speakers. The number of nodes in the DNN output layer is equal to the number of speakers.
Another feature is speaker+pass-phrase discriminant BN (SPK+phrase-BN) \cite{Yuan2015}, in which DNNs are trained to discriminate both speakers and pass-phrases simultaneously. This involves two loss functions: one for discriminating speakers and the other for discriminating pass-phrases. % Hence, there are two types of output nodes: one predicting the speakers and the other predicting pass-phrases. \htcomment{remove ``Hence, ...''?} 
The average of the two losses is used as the final criterion in the DNN multi-task learning procedure. We use the CNTK toolkit  \cite{yu2014introduction} for all BN-DNN training.

% \sscomment{``ASR contains context-dependent language model and our experiments were done just based on phoneme recognition. So, in my opinion, PHN-BN is more precise rather than BN-ASR to avoid misunderstanding...isn't it?'' }\\
% \ztcomment{That is true. The forced-alignment approach does contain language model though. As the labels are phones, it sounds good to use PHN-BN.} 

\subsection{Phone-discriminant BN features}

\begin{figure}[t]
   \centering\includegraphics[height=6.3cm,width=6.5cm]{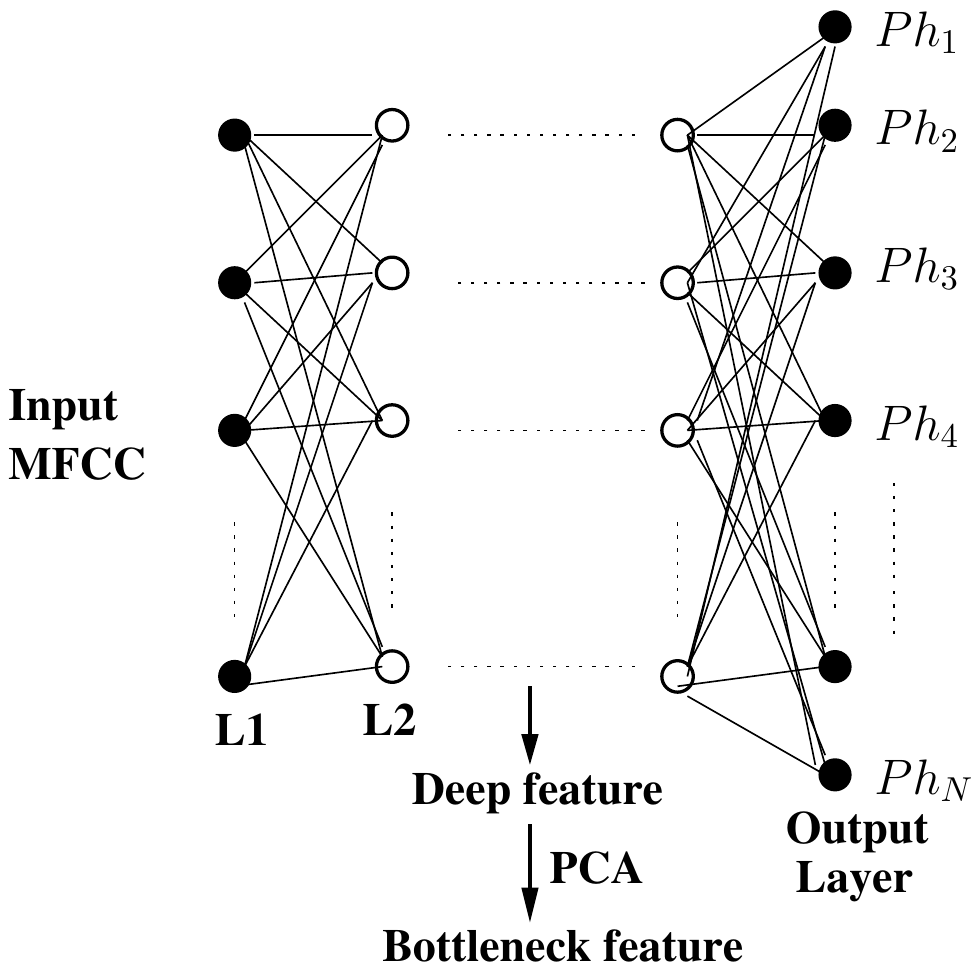}
   \caption{\it Bottleneck feature extraction from a DNN trained to discriminate phones.}
  \label{fig:sys1}
  \end{figure}

In the literature, triphone states or senones have been used as the BN-DNN target classes \cite{mclaren2015advances,Yuan2015}. This gives a large number of output neurons, e.g. 3001 tied-triphone-states in \cite{Yuan2015} and the performance is not promising. In this work, instead, we investigate the use of phones as the training target classes, which gives significantly lower class granularity. Specifically, DNNs are trained to discriminate phones and the number of nodes in the DNN output layer is equal to the number of phones as shown in Fig. \ref{fig:sys1}. % In this work, the resulting BN features are called phone-discriminant bottleneck features (PHN-BN). 
We consider three different speech recognizers for generating phone labels as detailed in the following. 

For PHN-BN1, the phoneme recognizer based on~\cite{Schwarz} is used to generate phoneme alignments for the RSR2015 database \cite{RSR2015}. % the Reddots challenge 2016 database  \ztcomment{It should be RSR2015 here. Changed}\akccomment{it's correct, RSR2015 only used for the training all DNN for BN i.e. ASR transcriptions for phone discriminate}. 
39 English phonemes are considered. The recognizer consists of three artificial neural network (ANNs) and each ANN has a single hidden layer with 500 neurons. A total of 23 coefficients are extracted as Mel-scale filter bank energies and the context of 31 frames are concatenated for long temporal analysis. This context is split into left and right blocks (with one frame overlap)~\cite{Schwarz}. Two front-end ANNs produce phoneme posterior probabilities for the two blocks separately, and the third back-end ANN merges the posterior probabilities from the two context ANNs.
% \\[1.2ex]

PHN-BN2 is based on an end-to-end segmental phoneme recognizer \cite{tang2017endtoend}. We use 40-dimensional log-Mel feature vectors as the input to the segmental model.
The segmental model consists of a 3-layer bidirectional long short-term memory (LSTM) with 256 cell units for each direction.
The segmental features are a combination of averaging over the hidden vectors of different parts of the segments and the length of the segment (termed FCB in \cite{tang2017sequence}).
The segmental model is trained on the TIMIT training set \cite{Timit} with the standard phone set including 47 phones and one label for silence.
The maximum phone duration is cap to 30 frames.
Marginal log loss \cite{tang2016endtoend} is optimized with Stochastic gradient descent (SGD) for 20 epochs with step size 0.1, gradient clipping of norm 5, and a batch size of one utterance.
The best model is chosen based on phone error rates from the first 20 epochs, and is trained for another 10 epochs in the same way except with the step size 0.75 decayed by 0.75 after each epoch.
We then decode using the best segmental model to obtain phone sequences for the utterances in the RSR2015 database \cite{RSR2015}. % the RedDots challenge 2016 database  \ztcomment{It should be RSR2015 here. Changed.}.

PHN-BN3 is based on forced alignments generated from the end-to-end segmental model \cite{tang2017endtoend}. Though trained end-to-end, the segmental model is able to produce excellent alignments without using any manual segmentation \cite{tang2016endtoend,tang2017sequence}. 

It is noted that in the ASR based approaches, 'sil' or 'pause' is included in the phoneme list for speech recognition or generating phone sequences. However, they are excluded from subsequently training DNNs that are used for BN feature extraction. In other words, a 'sil' or 'pause' model has the function of detecting the less energized or silence frames and then removing these frames from BN-DNN training. Table \ref{table:asr_phone} lists the phones available in different ASR systems, excluding 'sil' and 'pause'.

% However, evaluation data (target enrolment and test) of the SV considers only the  energy based VAD i.e. identical for all the systems. In other words,  only the DNNs (either using ASR transcriptions or speaker or speaker + pass-phrase label information) for the BN feature extraction is expected to influence prominently the overall SV performance. Indeed, it is more fair to comparison/understand the effect of different DNN based BN feature extraction for the TD-SV.  \ref{table:asr_phone} presents the list of phones are available in the different phone discriminating BN feature extraction.  \htcomment{Is the previous sentence true? I'm not using a VAD to remove frames while training and decoding.} \ztcomment{'sil' or 'pause' segments are not used for training DNNs. Reworded.} \akccomment{I added (for the BN feature extraction) to avoid the confusion, also added few more lines}
% at utterance level.

\begin{table}[H]
\caption{\it Lists of phones generated from different speech/phone recognition systems and used for training BN-DNNs.}
\begin{center}
\begin{tabular}{|c|l|}\cline{1-2}
System   & Phones \\ \hline 
PHN-BN1 & aa ae ah aw ay b ch d dh dx eh er ey\\
        &  f g hh ih iy jh k l m n  ng ow oy\\
        &  p r s sh t th uh uw v w y z  \\
        \hline
PHN-BN2 & aa ae ah ao aw ax ay b ch cl d dh dx \\
        & eh el en epi er ey f g hh ih ix iy jh\\
        & k l m n ng ow oy p r s sh t th  \\
        & uh uw v vcl w y z zh    \\ \hline 
PHN-BN3 & aa ae ah ao aw ay b ch d dh eh er ey\\
        & f g hh ih iy jh k l m n ng ow oy\\
        & p r s sh t th uh uw v w y z zh   \\ \hline         
\end{tabular}
\end{center}
\label{table:asr_phone}
\end{table}

\subsection{Time-contrastive learning based BN features}

\begin{figure}[H]
  \centering\includegraphics[height=4.0cm,width=9.0cm]{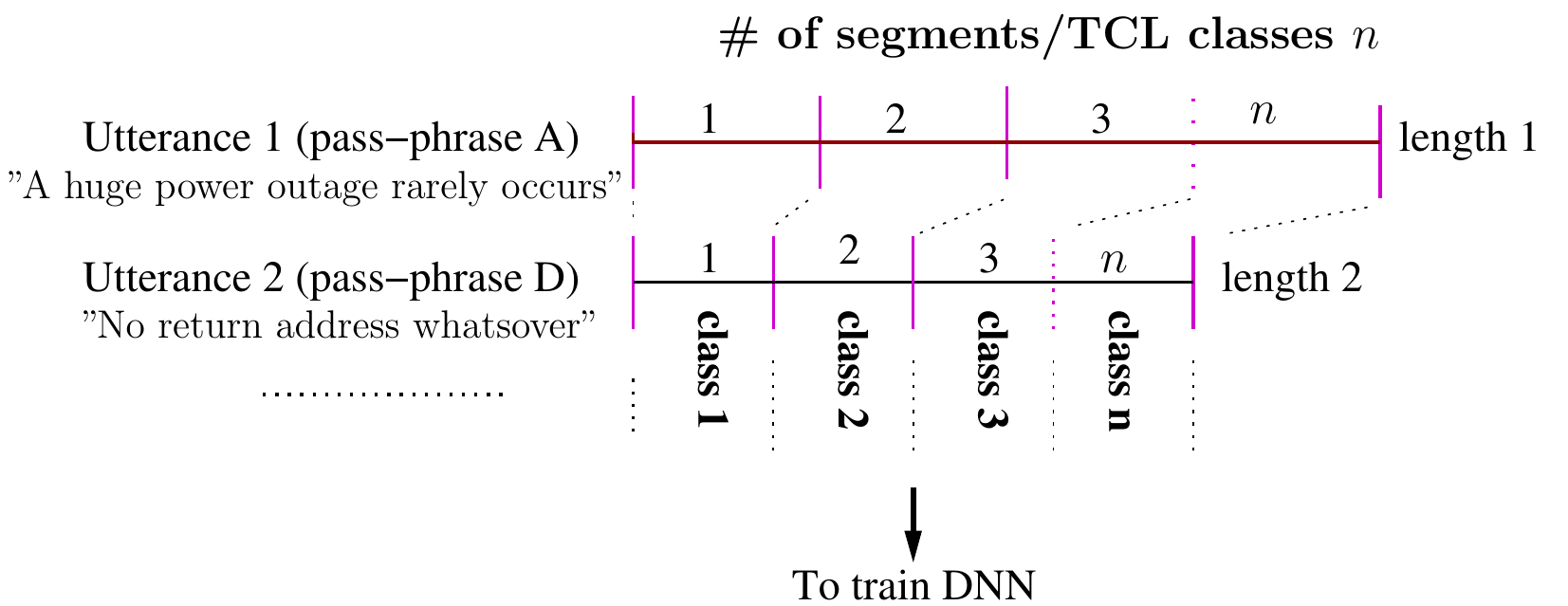}
    \caption{\it Segmentation of speech utterances for BN feature extraction in uTCL.}
   \label{fig:utcl}
 \end{figure}

We recently proposed to apply TCL to extract BN features for TD-SV \cite{AchintyaNips2017}. There are two ways to implement the TCL method. One is utterance-wise TCL (uTCL), in which each utterance for training DNNs is uniformly divided into $N$ segments. The number of segments $N$ is equal to the number of classes $N$ in TCL, i.e., the number of  output nodes in DNNs. Speech frames within a particular segment are assigned a class label as follows:
\begin{equation}
{\footnotesize
  \underbrace{(x_1, ..., x_M)}_\text{Class $1$}, \ldots, \underbrace{(x_{(n-1)M+1}, ..., x_{nM})}_\text{Class $n$}, \ldots, \underbrace{(x_{(N-1)M+1}, ..., x_{NM})}_\text{Class $N$} 
  }
   \end{equation}
where $n$ and $M$ indicate the segment index (as well as the class ID) and the number of frames within a segment, respectively. 
Afterwards, DNNs are trained to discriminate the frames among the classes. We vary the value of $N$ in order to study the effect of different numbers of classes in TCL on TD-SV. Fig.\ref{fig:utcl}  illustrates the segmentation of speech utterances for BN feature extraction in uTCL. 

The other way of realizing TCL for speech is called stream-wise TCL (sTCL) \cite{AchintyaNips2017}. It is similarly to uTCL, with the only difference being that  training data of the DNNs are first randomly concatenated into a single speech stream. The single speech stream is then partitioned into segments of $6$ frames each (chunk). While uTCL attempts to capture the structures in a speech corpus, e.g. repeating sentences, sTCL constructs DNN training in much higher degree of randomness. 
% The objective of the sTCL segmentation is to capture short-time speech events in the speech signal instead of temporal structure like in uTCL and see whether it is useful for the TD-SV. 
% \begin{figure}[H]
%  \centering\includegraphics[height=3.4cm,width=8.2cm]{sTCL.eps}
%  \caption{\it sTCL based segmentation and define class label for  bottleneck feature extraction.}
% \label{fig:stcl}
% \end{figure}
% For more detail about the TCL concept see \cite{Aapo2016,AchintyaNips2017}. % \\[1.2ex]

To obtain BN features in the respective systems, the output of a DNN hidden layer at frame-level is projected  onto a lower dimensional space by using principle component analysis (PCA). % \emph{called bottleneck features}.

\section{Segment-based Clustering}

%\section{Time-contrastive learning based BN features and segment based clustering}
%In this section, we present TCL based bottleneck features and a segment-based clustering method that re-assigns labels to segments. 
% \subsection{Segment based clustering}

As segment classes in TCL are defined or assigned by uniformly segmenting speech signals in unsupervised manner, segment contents in each class are inevitably heterogeneous. This motivates us to devise a clustering algorithm to group similar speech segments together and form new groups/classes. This is expected to be beneficial for DNN training, thus leading to improved BN features. In this section, we propose a segment-based clustering method, which re-assigns labels to segments, as follows.
\begin{itemize}
\item[] {\bf Step1:} Pool together all speech segments belonging to a particular class $c_n$ and derive the class specific GMM, $\lambda_{n}$, from the GMM-UBM (trained on the TIMIT dataset) through maximum a posteriori (MAP) adaptation.
\item[] {\bf Step2:} Classify each speech segment using newly-derived class-specific GMMs based on the maximum likelihood approach,
\begin{eqnarray}
        \hat{i}= \mbox{arg }\max_{1\leq i \leq N} p(S_j|\lambda_{i}) \label{ML_GMM}
\end{eqnarray}
where $S_j$ denotes the set of feature vectors in the $j^{th}$ speech segment.
\item[] {\bf Step3:} Check whether the stop criteria are met. If yes, go to next step. Otherwise, go to \emph{Step 1} and repeat the process. % Derive a new set of GMMs by pooling together all speech segment and using the labels generated from the classification \emph{Step 2} 
% \item[] {\bf Step4:} Repeat \emph{Step2} to \emph{Step3} for a few iterations
\item[] {\bf Step 4:} Output the new class labels for speech segments (for training the BN-DNN) 
\end{itemize}

\begin{figure}[H]
\centering\includegraphics[height=4.0cm,width=8.2cm]{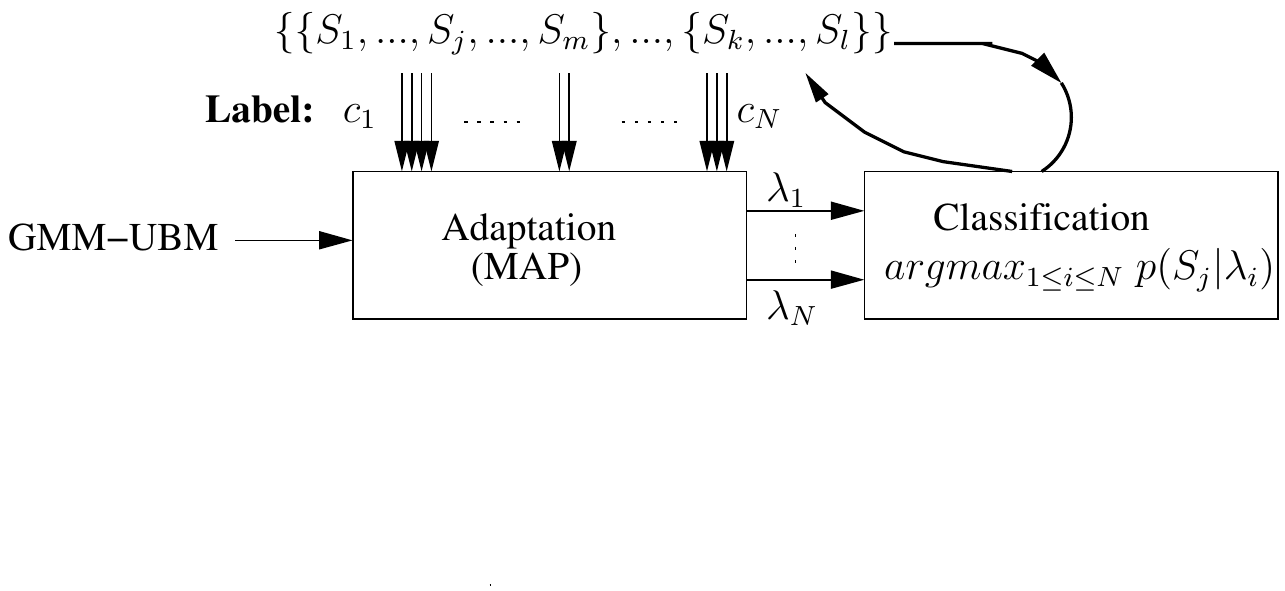}
\caption{\it Illustration of segment-based clustering for speech data with $N$ classes.}
\label{fig:clustering}
\end{figure}

Fig. \ref{fig:clustering}  illustrates the  clustering method. In this work, the method is used in combination with TCL-BN and PHN-BN. 
In the experiment of this work, the stop criterion is that \emph{Step1} and \emph{Step2} are repeated $5$ iterations, which is found to give a stable set of clusters, i.e. the clusters do not change much. This choice is for simplicity and computational time efficiency. 
While the proposed algorithm is for clustering, it differs from the conventional K-means algorithm \cite{MacQueen67} by being based on probability than Euclidean distance. It also differs from the expectation-maximization (EM) algorithm for training GMMs \cite{dempster77}. First, it is for clustering than density estimation. Secondly, it is based on segments rather than single frames. Thirdly, cluster-specific GMMs are updated from the GMM-UBM (a priori distribution) through MAP adaptation in contrast to the maximization step in the EM algorithm where cluster-specific Gaussian models are directly calculated on the data belonging to each cluster.

The way the proposed clustering method iteratively increases the likelihood of segments shares some similarity to the generation of forced alignment in ASR training \cite{young2006htk} where triphone segments are gradually refined through an align-realign process. There are also a number of differences between them as follows: 1) forced alignment is generated by using a given text transcription (without time stamps) while the segment clustering method does not use any transcription, 2) the forced alignment sequence is fixed by the text while segments have no fixed ordering in the segment clustering, 3) segment durations of forced alignment change during the iterative process while they are fixed for the segment clustering, and 4) hidden Markov models or hybrid models are used for forced alignment while GMMs are used for the segment clustering method.

\section{Speaker Verification Methods}
We consider two best-known methods for speaker verification: GMM-UBM and i-vector.
 \subsection{The GMM-UBM method} 
\begin{figure}[h]
   \centering\includegraphics[height=6.0cm,width=9.0cm]{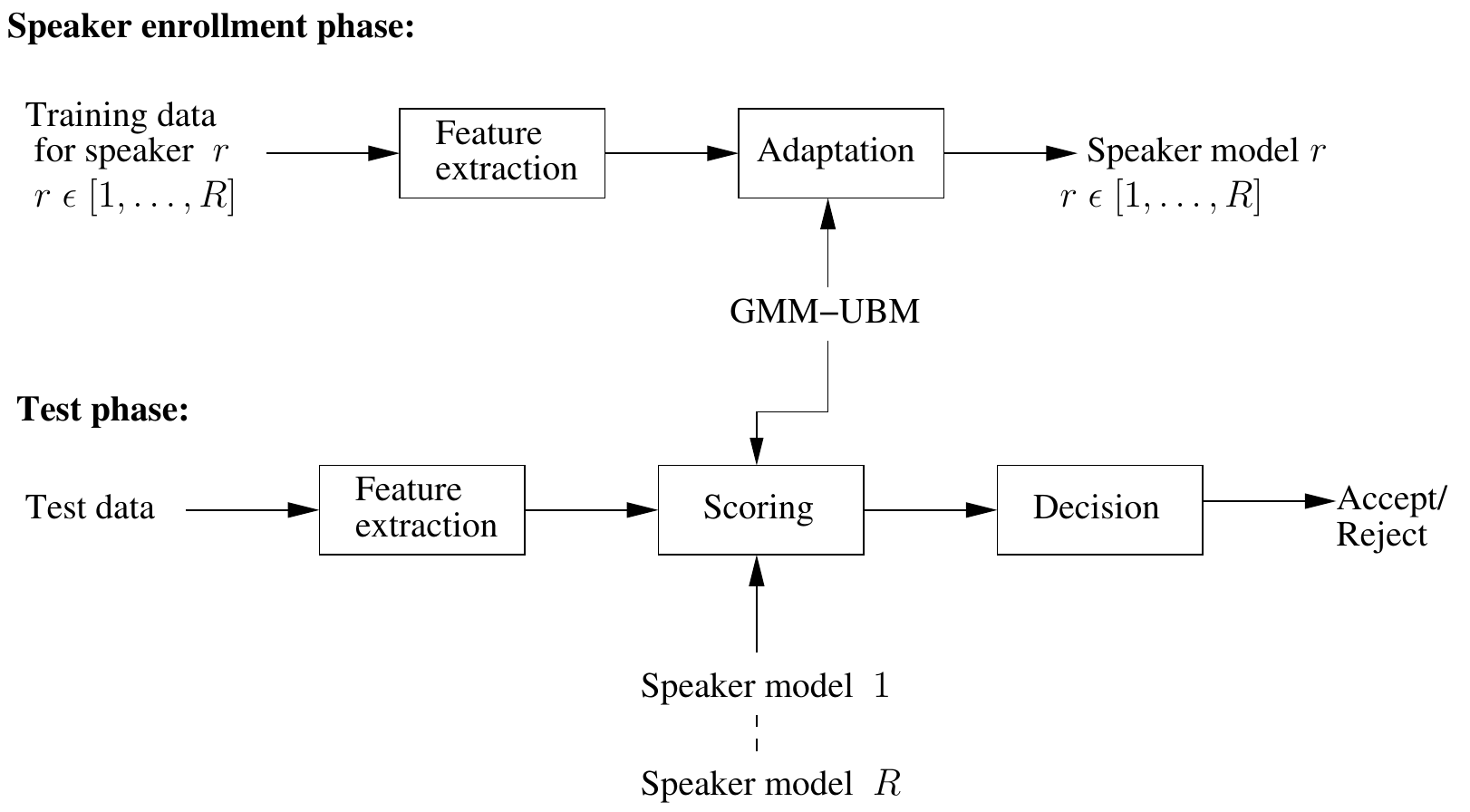}
   \caption{\it GMM-UBM based speaker verification.}
  \label{fig:GMM_UBM}
  \end{figure}

 As per \cite{reynold00}, a target speaker model is derived from GMM-UBM with MAP adaptation using the training data of the target speaker during the enrolment phase as illustrated in Fig.\ref{fig:GMM_UBM}. 
 
During the test phase, the feature vectors of a test utterance $Y=\{y_1,y_2, \ldots, y_T\}$ is scored against the claimant model (i.e. the target speaker model)  $\lambda_r$ and GMM-UBM $\lambda_{ubm}$. Finally, the  log likelihood ratio (LLR)  value is calculated using the scores between the two models
\begin{equation}
LLR(Y) = \frac{1}{T} \sum_{t=1}^{T}\{\log \;p(y_t|\lambda_r) - \log\; p(y_t|\lambda_{ubm})\}
\end{equation}
%$\textbf{y}_t$ denotes the $t^{th}$ frame/feature vector.

It is well established \cite{Yuan2015,Delgado2016Asru} that GMM-UBM performs better than i-vector   for speaker verification using short speech utterances.

\subsection{The i-vector method} 
In this framework, a speech utterance is represented by a vector called i-vector \cite{Deka_ieee2011}.
The i-vector $w$ is obtained by projecting the speech utterance onto a subspace $T$ (called total variability space or T-matrix) of a GMM-UBM super-vector, where speaker and channel information is dense. It is generally expressed as,
\begin{equation}
M = m + T w
\end{equation}
where $w$ is an i-vector, $M$ and $m$ denote the utterance dependent GMM super-vector, the speaker-independent GMM super-vector obtained by concatenating the mean vectors from the GMM-UBM, respectively, and $T$ the total variability space. For more details refer to \cite{Deka_ieee2011}.

\begin{figure}[H]
\centering\includegraphics[height=5.0cm,width=9.0cm]{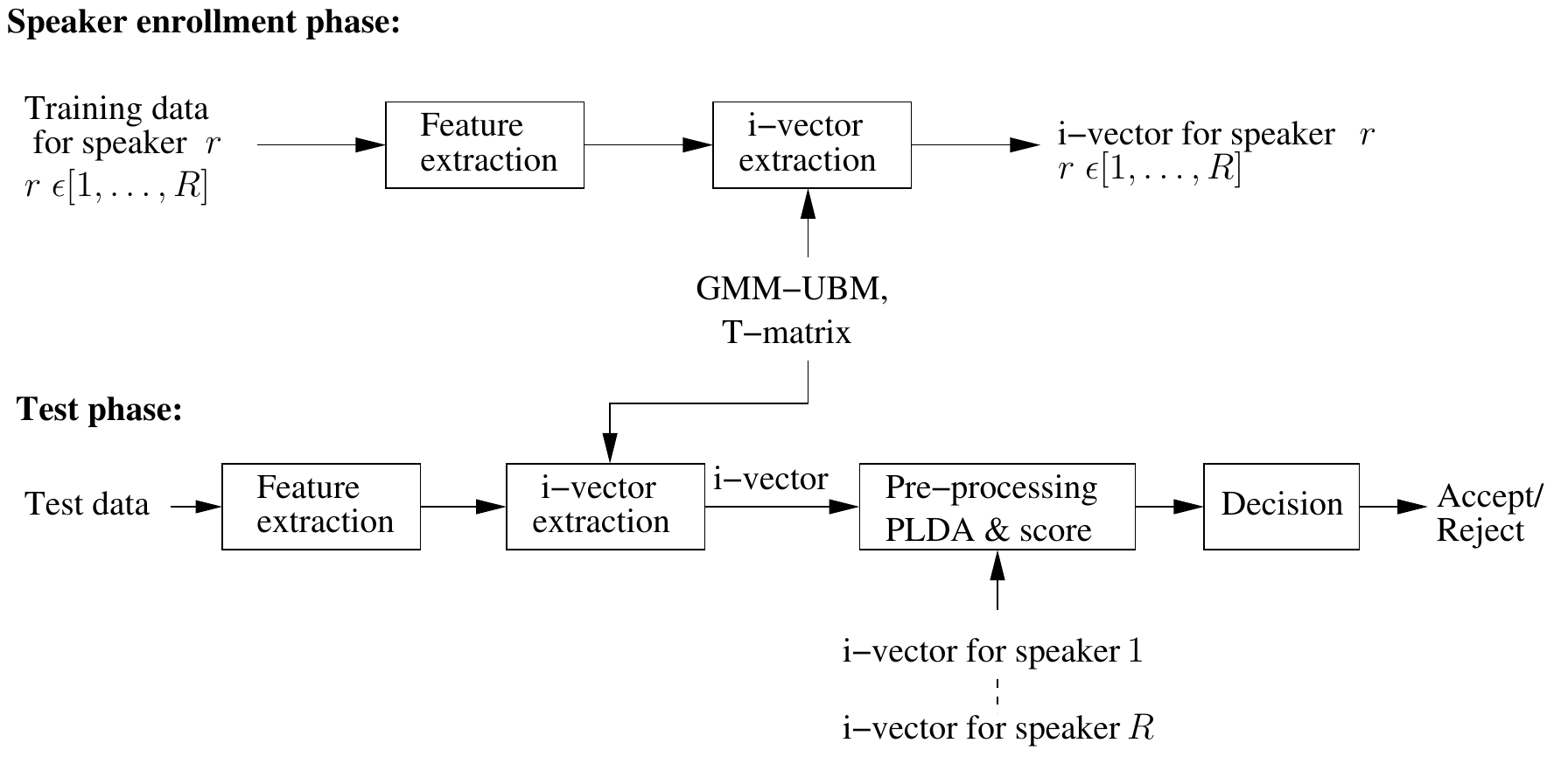}
   \caption{\it Illustration of i-vector based speaker verification.}
  \label{fig:ivector}
  \end{figure}

During the enrolment, each target is represented by an average i-vector computed over his/her training utterance-wise (or speech session-wise) i-vectors. In the test phase, the score between the i-vector of a test utterance and the claimant specific i-vector (obtained during enrolment) is calculated using probability linear discriminate analysis (PLDA). Fig.\ref{fig:ivector} illustrates the speaker enrolment and test phases of i-vector based speaker verification.

PLDA represents an i-vector in the joint factor analysis (JFA) framework as
\begin{eqnarray}
w=\mu_w+\Phi y + \Gamma z+\epsilon \label{eq:plda1}
\end{eqnarray}
where $\Phi$ and $\Gamma$ are  matrices denoting the \emph{eigen voice} and \emph{eigen channel} subspaces, respectively. $y$ and $z$ are the speaker and channel factors, respectively, with a priori normal distribution. $\epsilon$ represents the residual noise. $\Phi$, $\Gamma$ and $\epsilon$ are  iteratively updated  during the training process by pooling together a numbers of  i-vectors per speaker class from many speakers. 
During  test, the score between two i-vectors ($w_1$, $w_2$) is calculated as:
\begin{eqnarray}
score(w_1,w_2)=\log\frac{p(w_1,w_2|\theta_{tar})}{p(w_1,w_1|\theta_{non})} \label{eq:pldascore}
\end{eqnarray}
where hypothesis $\theta_{tar}$ states that $w_1$ and $w_2$ come from the same speaker, and  hypothesis $\theta_{non}$ states that they are from different speakers. For more details about the PLDA based scoring see~\cite{Pierre-interspeech2012,Prince2012,SenoussaouiInterspch2011}. Before scoring, i-vectors are conditioned to reduce the session variability with two iterations of  spherical normalization (sph) as in \cite{Pierre-interspeech2012}.

\section{Experimental Set-ups}
Experiments were conducted on the 'm-part-01' task (male speakers) of the RedDots database as per protocol  \cite{RedDots}.
There are $320$ pass-phrase dependent target models for training. Each target has three speech files for training. Each utterance is very short in duration (approximately 2-3s in duration). Three types of non-target trials are available  for the evaluation of text dependent speaker verification system. Table \ref{table:trial_info} presents the number of different trial available in evaluation.\\  
{\bf True-trials:} when a target speaker claims by pronouncing the same pass-phrase as enrolment in the testing phase.\\
{\bf Target-wrong (TW):}  when a target speaker claims by pronouncing a different pass-phrase in the testing phase.\\
{\bf Imposter-correct (IC):} when an imposter speaker claims by speaking the same pass-phrase as target in the enrolment phase.\\
{\bf Imposter-wrong (IW):} when an imposter speaker claims by speaking a wrong pass-phrase.\\

\begin{table}[H]
\caption{\it Numbers of different trials available for the TD-SV evaluation on the RedDots database.}
\begin{center}
\begin{tabular}{|l|c|c|c|}\cline{1-4}
\# of     & \multicolumn{3}{c|}{\# of non-target trials} \\ \cline{2-4}
True   & Target  & Imposter & Imposter \\
trials    &-wrong   &-correct & -wrong\\ \hline 
3242      & 29178   &120086   & 1080774 \\ \hline
\end{tabular}
\end{center}
\label{table:trial_info}
\end{table}

\begin{table*}[t]
%\scriptsize{
\caption{\it TD-SV results of MFCCs and BN features on the m-part-01 task of the RedDots database using the GMM-UBM method. Gray-colored text shows the results of BN features extracted from a non-default hidden layer to provide further insights about the behavior of the corresponding BN extraction methods, while those features will not be  used in real systems. }
\begin{center}
\begin{tabular}{|lcccccc|c|}\cline{1-8}
 Feature     & DNN    & \# of  & Clustering &\multicolumn{3}{c}{Non-target type [\%EER/(minDCF$\times$ 100)]} & Average\\ 
             &Lyr.    & classes& without: $\times$           &Target-          & Impostor-        & Impostor-     & (EER \\
              &       &        &  with: \checkmark  & wrong        & correct      & wrong               &  /minDCF)   \\  \hline \hline  
 MFCC         &       &        & -          & 5.12/2.17    & 3.33/1.40    & 1.14/0.47   & 3.19/1.35 \\ 
              &       &        &           &              &              &             &     \\ 
 SPK-BN      & \textcolor{gray}{L2} & \textcolor{gray}{300}   & - & \textcolor{gray}{4.81/1.66}    & \textcolor{gray}{3.28/1.39}    & \textcolor{gray}{1.29/0.43}   & \textcolor{gray}{3.13/1.16} \\
              &  L4 &           & -         & 4.59/1.65    & 3.05/1.35    &1.11/0.38    & {\bf2.91/1.13}\\
              &     &           &           &   &              &             &     \\ 
SPK+phrase-BN      & \textcolor{gray}{L2}  & \textcolor{gray}{327}   & - &  \textcolor{gray}{4.79/1.66}    & \textcolor{gray}{3.20/1.40}    & \textcolor{gray}{1.30/0.42}   & \textcolor{gray}{3.10/1.16} \\
       & L4  &                 &    -           & 4.53/1.64    &3.07/1.34     &1.17/0.38    & {\bf2.92/1.12} \\  
              &      &         &              &              &              &             &     \\
PHN-BN1       & L2  & 38       &   $\times$   & 2.31/0.71    & 3.14/1.29    & 0.61/0.20   & {\bf 2.02/0.73} \\  
              & \textcolor{gray}{L4}  &  &  $\times$     & \textcolor{gray}{7.77/4.07}    & \textcolor{gray}{6.53/3.41}    & \textcolor{gray}{3.14/1.47}   & \textcolor{gray}{5.81/2.98}  \\
              %&     &       &              &              &             &           \\ 
              & L2  &     &     \checkmark & 2.32/0.74    & {\bf 2.96/1.22}& 0.64/0.18 &{\bf 1.97/0.72} \\          
               & \textcolor{gray}{L4}&  & \checkmark     & \textcolor{gray}{3.67/1.65}    & \textcolor{gray}{5.24/2.56}    & \textcolor{gray}{1.32/0.48}   & \textcolor{gray}{3.41/1.58} \\ 
               &     &              &              &             &         &   \\
 PHN-BN2       & L2  & 47    & $\times$ & 2.25/0.78    & 2.89/1.30    & 0.61/0.22   & {\bf1.92/0.77}  \\
              & \textcolor{gray}{L4}  &       & $\times$ & \textcolor{gray}{2.29/0.86}    & \textcolor{gray}{4.99/2.33}    & \textcolor{gray}{0.80/0.33}   & \textcolor{gray}{2.69/1.17}  \\
             % &     &       &              &              &             &    \\
 & L2  &  & \checkmark & 2.14/0.79    & 2.68/1.21    & 0.61/0.22   & {\bf1.81/0.74} \\
                   & \textcolor{gray}{L4} &  & \checkmark & \textcolor{gray}{2.71/1.13}     & \textcolor{gray}{4.04/1.82}    & \textcolor{gray}{0.95/0.34}   & \textcolor{gray}{2.57/1.10} \\
              &     &              &              &             &         &   \\

  PHN-BN3       & L2  & 39    &   $\times$ & 1.79/0.72    & 3.08/1.41    &  0.55/0.15  & {\bf1.81/0.76} \\     
               (ASR force-alignment)   & \textcolor{gray}{L4}  &       & $\times$&  \textcolor{gray}{1.70/0.65}    & \textcolor{gray}{4.75/2.46}    & \textcolor{gray}{0.74/0.21}   &  \textcolor{gray}{2.39/1.11}  \\ 
                %&     &              &              &             &         &   \\
               & L2  &  &  \checkmark  & 2.08/0.70    & {\bf2.83}/1.18& 0.55/0.18  & 1.82/{\bf0.69}    \\
                     & \textcolor{gray}{L4}  &  & \checkmark & \textcolor{gray}{2.89/1.18}    & \textcolor{gray}{4.56/2.18}     & \textcolor{gray}{1.17/0.4}2  & \textcolor{gray}{2.89/1.26}     \\
                     &     &   &  &           &               &            &               \\ 
  sTCL-BN            & L2 &10 &   $\times$    &4.42/1.61       &  3.08/1.32    & 1.12/0.38  & {\bf2.88/1.10} \\
                     & \textcolor{gray}{L4} &  & $\times$ &\textcolor{gray}{4.68/1.68}        &  \textcolor{gray}{3.23/1.39}      & \textcolor{gray}{1.23/0.40}     & \textcolor{gray}{3.05/1.16} \\
%                %15    & 4.33/1.66  &  3.02/1.38    & 1.14/0.39  & 2.83/1.14 \\
                & L2  &   & \checkmark &2.83/1.03&  2.86/1.34 & 0.98/0.26 & {\bf 2.23/0.87}  \\
                & \textcolor{gray}{L4}  &   & \checkmark & \textcolor{gray}{9.57/6.26}        & \textcolor{gray}{7.80/4.06}        & \textcolor{gray}{3.89/2.37}     & \textcolor{gray}{7.09/4.23} \\
 %&   &  15     & 3.33/1.07        &  3.20/1.37      & 0.98/0.31     & {\bf2.50/0.92} \\
                &     &       &    &            &              &            &          \\
  uTCL-BN        & L2  & 10    &   $\times$ & 1.88/0.65     &  3.14/1.44   & 0.64/0.19  & 1.89/0.76 \\
                 & \textcolor{gray}{L4}  &   & $\times$       & \textcolor{gray}{19.63/9.95}       &\textcolor{gray}{18.51/8.93}       &\textcolor{gray}{11.69/6.89}     & \textcolor{gray}{16.61/8.59} \\
       & L2    &  &  \checkmark & 1.91/0.60     &  {\bf2.77/1.17}& 0.70/0.18& {\bf1.79/0.65} \\ 
               & \textcolor{gray}{L4}    &  & \checkmark & \textcolor{gray}{5.98/3.61}        & \textcolor{gray}{7.44/3.91}       & \textcolor{gray}{2.52/1.35}     & \textcolor{gray}{5.31/2.96} \\ \hline
\end{tabular} 
\end{center}
\label{table:table1}
 \end{table*}

\begin{table*}[htp!]
\caption{TD-SV results of TCL-BN features with/without clustering on the m-part-01 task of the RedDots database using the GMM-UBM method. The average percentage of EER and $MinDCF \times 100$ for MFCC are ${\bf3.19}$ and ${\bf1.35}$, respectively.}
\begin{subtable}[t]{.5\textwidth}
\scriptsize{
\subcaption{sTCL } 
\vspace*{-0.3cm}
\raggedright
\begin{center}
\setlength\tabcolsep{2.5pt}
\begin{tabular}{|lccccc|c|}\cline{1-7}
Feature  &DNN   &TCL         & \multicolumn{3}{c}{Non-target type [\%EER/(MinDCF$\times$ 100)]} &  \\ 
        &Lyr.  &classes             & Target-      & Impostor-        & Impostor-     & Average \\ 
     &      &({\bf $N$})         &wrong            & correct          & wrong         & EER/MinDCF   \\ \hline  \hline 
sTCL & L2   & 2                  & 4.50/1.69       & 3.12/1.39        & 1.01/0.39     & 2.88/1.16  \\
     &      & 3                  & 4.60/1.67       & 3.13/1.40        & 1.20/0.40     & 2.98/1.16  \\
     &      & 4                  & 4.57/1.65       & 3.14/1.38        & 1.17/0.40     & 2.96/1.14 \\
     &      & 5                  & 4.53/1.65       &  3.16/1.39      & 1.06/0.40      & 2.91/1.15 \\
     &      & 6                  & 4.38/1.64       & 3.14/1.37        & 1.07/0.39     & {\bf2.86}/1.13 \\
     &      & 7                  & 4.62/1.69       &  3.10/1.34       & 1.29/0.41     & 3.00/1.15 \\
     &      & 8                  & 4.44/1.63       & 3.17/1.39        & 1.11/0.40     & 2.90/1.14 \\
     &      & \cellcolor[gray]{0.9}10 & \cellcolor[gray]{0.9}4.42/1.61  &  \cellcolor[gray]{0.9}3.08/1.32       & \cellcolor[gray]{0.9}1.12/0.38     & \cellcolor[gray]{0.9}{\bf2.88/1.10} \\
     &      & 12                 &  4.50/1.66       &  3.14/1.41       & 1.14/0.41     & 2.93/1.16  \\
     &      & 15                 & 4.33/1.66        &  3.02/1.38      & 1.14/0.39     & {\bf2.83}/1.14 \\
     &      & 20                 & 4.35/1.66        &  3.10/1.38      & 1.14/0.39     & 2.86/1.14  \\
     &      & 40                 & 4.38/1.65        &  3.17/1.38      & 1.15/0.39     & 2.90/1.14  \\
     &      &                     &                   &                  &                &     \\
     & L4   & 2                  & 4.48/1.58         & 3.20/1.32       & 1.17/0.40     & 2.95/1.10  \\
     &      & 3                  & 4.44/1.64         & 3.36/1.38       & 1.29/0.42     & 3.03/1.15  \\
     &      & 4                  & 4.65/1.65         & 3.23/1.38       & 1.17/0.40     & 3.02/1.14 \\
     &      & 5                  & 4.52/1.67        &  3.08/1.40      & 1.23/0.39     & 2.94/1.15 \\
     &      & 6                  & 4.50/1.63         & 3.23/1.36       & 1.24/0.40     & 2.99/1.13 \\
     &      & 7                  &  4.45/1.67        &  3.02/1.33      & 1.11/0.39     & 2.90/1.13 \\
     &      & 8                  &  4.65/1.66        &  3.20/1.38      & 1.04/0.40     & 2.97/1.15 \\
     &      & 10                  & 4.68/1.68        &  3.23/1.39      & 1.23/0.40     & 3.05/1.16 \\
     &      & 12                  &  4.50/1.65       &  3.14/1.38      & 1.26/0.38     & 2.97/1.14  \\
     &      & 15                  & 4.44/1.73        &  3.11/1.38      & 1.20/0.39     & 2.92/1.17 \\
     &      & 20                  & 4.47/1.67        &  3.20/1.38      & 1.13/0.40     & 2.93/1.15 \\ 
     &      & 40                 &  4.59/1.72        &  3.17/1.40      & 1.23/0.41     & 3.00/1.17  \\
     &      &                     &                   &                  &                &     \\
\bf{+clustering} & L2     & 2    &  2.99/0.99       &   3.08/1.40        & 0.99/0.25   & 2.35/0.88  \\
                 &        & 3    & 2.80/0.97        & 3.00/1.43         & 0.83/0.27    & 2.21/0.89  \\
                 &        & 4    & 4.34/1.66        & 3.17/1.39         & 1.32/0.39    & 2.95/1.15 \\
                 &        &5     & 3.39/1.16        &  3.36/1.44       & 1.07/0.32     & 2.61/0.97 \\
                 &        & 6    & 3.32/1.14        & 3.23/1.46        & 1.05/0.31     & 2.53/0.97  \\
                 &        & 7    & 4.44/1.67        &  3.17/1.27       & 1.41/0.39     & 3.01/1.14  \\
                 &        & 8    & 3.14/1.17        &     3.05/1.40    &  0.95/0.33    & 2.38/0.97\\
                 &        & \cellcolor[gray]{0.9}10 & \cellcolor[gray]{0.9}2.83/1.03        &   \cellcolor[gray]{0.9}2.86/1.34   & \cellcolor[gray]{0.9}0.98/0.26 & \cellcolor[gray]{0.9}{\bf 2.23/0.87} \\
                 &        & 12   &  3.14/1.10       & 3.11/1.41        & 1.02/0.31     & 2.43/0.94  \\
                 &        & 15   & 3.33/1.07   &  3.20/1.37       & 0.98/0.31     &  2.50/0.92 \\
                 &        & 20   & 3.05/1.13        &  2.93/1.36       & 0.92/0.30     & 2.30/0.93 \\
                 &        & 40   & 4.25/1.58        &  3.17/1.37       & 1.07/0.36     & 2.83/1.11  \\ 
                 &      &                     &                   &                  &                &     \\
                 & L4     & 2    & 11.25/5.72       & 12.02/5.89       & 7.18/3.10     & 10.15/4.90 \\
                 &        & 3    & 18.07/9.92       & 17.98/8.34       & 12.46/6.21    & 16.15/8.16\\
                 &        & 4    & 4.38/1.69        & 3.10/1.36        & 1.20/0.42     & 2.89/1.16\\
                 &        & 5    &  18.53/9.34      & 17.76/7.64       & 14.15/5.50    & 16.82/7.50 \\
                 &        & 6    & 15.39/9.33       & 14.12/6.55       & 9.74/4.65     & 13.08/6.84 \\
                 &        & 7    & 4.59/1.67        & 3.08/1.26        & 1.41/0.42     & 3.03/1.12\\
                 &        & 8    & 11.53/7.33       & 10.05/5.06       & 5.71/2.87     & 9.10/5.09 \\
                 &        & 10   & 9.57/6.26        & 7.80/4.06        & 3.89/2.37     & 7.09/4.23 \\
                 &        & 12   & 7.75/4.05        & 6.72/3.29        & 2.81/1.53     & 5.76/2.96 \\
                 &        & 15   & 7.74/4.50        & 6.05/3.15       & 2.84/1.64     & 5.54/3.10   \\
                 &        & 20   & 6.90/3.62        & 6.14/2.92       & 2.93/1.39     & 5.32/2.64  \\
                 &        & 40   & 4.82/1.79        & 3.57/1.45        & 1.29/0.46     & 3.23/1.23\\
                      \hline 
\end{tabular}
\end{center}
\label{table:sTCL}
} 
\end{subtable}
\begin{subtable}[t]{.5\textwidth}
\raggedleft
\scriptsize{
\subcaption{uTCL}
\vspace*{-0.3cm}
\begin{center}
\setlength\tabcolsep{2.5pt}
\begin{tabular}{|lccccc|c|}\cline{1-7}
Feature  &DNN   &TCL         & \multicolumn{3}{c}{Non-target type [\%EER/(MinDCF$\times$ 100)]} &  \\ 
        &Lyr.  &classes             & Target-      & Impostor- & Impostor- & Average \\ 
     &      &({\bf $N$})         &wrong         & correct   & wrong     & EER/MinDCF        \\ \hline  \hline
uTCL & L2   & 2                   & 2.12/0.71       & 3.28/1.48        & 0.70/0.22     & 2.03/0.80  \\
     &      & 3                   & 2.00/0.73       & 3.43/1.50        & 0.77/0.21     & 2.07/0.81  \\
     &      & 4                   & 2.06/0.73       & 3.20/1.51        & 0.78/0.21     & 2.02/0.81\\
     &      & 5                   & 2.05/0.64        &  3.30/1.51      & 0.58/0.21     & 1.98/0.79 \\
     &      & 6                   & 2.39/0.88        & 3.39/1.54       & 0.74/0.28     & 2.17/0.90 \\
     &      & 7                   & 4.75/1.66        & 3.33/1.38       & 1.43/0.43     & 3.17/1.16  \\
     &      & 8                   & 2.59/1.02        & 3.60/1.63       & 0.92/0.35     & 2.37/1.00  \\
     &      & \cellcolor[gray]{0.9}10             & \cellcolor[gray]{0.9}1.88/0.65        &  \cellcolor[gray]{0.9}3.14/1.44      & \cellcolor[gray]{0.9}0.64/0.19     & \cellcolor[gray]{0.9}{\bf1.89/0.76} \\
     &      & 12                  & 1.88/0.64        &  3.39/1.54      & 0.80/0.211    &  2.02/0.80 \\
     &      & 15                  & 4.47/1.62        &  3.14/1.38      & 1.26/0.37     & 2.96/1.13 \\
     &      & 20                  & 4.38/1.59        &  3.13/1.33      & 1.35/0.38     & 2.95/1.10 \\ 
     &      & 40                  & 4.56/1.67        &  3.11/1.38      & 1/32/0.41     & 3.00/1.15 \\
     &      &                     &                   &                  &                &     \\
     & L4   & 2                   & 13.73/8.33       & 13.64/6.60      & 8.06/4.23     & 11.81/6.39  \\
     &      & 3                   & 19.82/9.96       & 17.63/9.93      & 11.25/8.01    & 16.23/9.30  \\
     &      & 4                   &  22.29/9.99      & 19.74/9.97      & 13.97/9.83    & 18.66/9.93 \\
     &      & 5                   & 15.79/9.98       &13.69/8.73       &8.18/6.61      & 12.55/8.44\\
     &      & 6                   & 11.66/7.90       &10.71/5.67       &5.53/3.33      & 9.30/5.63\\
     &      & 7                   & 4.62/1.63        & 3.14/1.36       &1.07/0.41      & 2.95/1.13  \\
     &      & 8                   & 10.17/7.40       & 9.50/5.40       & 4.44/2.88     & 8.04/5.22  \\
     &      &10                   & 19.63/9.95       &18.51/8.93       &11.69/6.89     & 16.61/8.59 \\
     &      & 12                  & 16.77/9.96       & 15.94/8.52      &8.90/6.08      & 13.87/8.19  \\
     &      &15                   & 4.43/1.62        & 3.17/1.32       & 1.14/0.38     & 2.91/1.11 \\
     &      & 20                  & 4.62/1.63        & 3.10/1.34       & 1.29/0.39     & 3.00/1.12 \\
     &      & 40                  & 4.41/1.65        & 3.11/1.37       & 1.13/0.38     & 2.88/1.13 \\
     &      &                     &                   &                  &                &     \\
\bf{+clustering} & L2   & 2       &  2.37/0.73      & 3.07/1.31        & 0.69/0.24     & 2.04/0.76  \\  
                 &      & 3       & 4.50/1.62       & 3.11/1.35        & 1.20/0.36     & 2.94/1.11  \\
                 &      & 4       & 4.41/1.62       & 3.05/1.36        & 1.41/0.39     & 2.96/1.12  \\
                 &      &  5      & 2.06/0.69        &  2.94/1.30      & 0.70/0.20     &  1.90/0.73 \\
                 &      & 6       & 2.25/0.72       & 2.99/1.32        & 0.82/0.24     &  2.02/0.76 \\
                 &      & 7       & 2.12/0.74        & 2.89/1.28       & 0.74/0.23     &  1.92/0.75 \\
                 &      & 8       & 1.99/0.65        & 2.74/1.26       & 0.61/0.19     & 1.78/0.70  \\
                 &      &\cellcolor[gray]{0.9} 10 & \cellcolor[gray]{0.9}1.91/0.60        &  \cellcolor[gray]{0.9}2.77/1.17      & \cellcolor[gray]{0.9}0.70/0.18     & \cellcolor[gray]{0.9}{\bf1.79/0.65} \\
                 &      & 12      & 1.94/0.63        &  2.74/1.19      &  0.58/0.17    & 1.75/0.66  \\
                 &      & 15      & 1.88/0.59        &  2.81/1.23      & 0.67/0.16     & 1.79/0.66 \\
                 &      & 20      & 2.25/0.75        &  2.74/1.25      & 0.64/0.19     & 1.88/0.73 \\ 
                 &      & 40      & 2.73/0.93        &  2.83/1.31      & 0.89/0.25     & 2.15/0.83  \\ 
                 &      &                     &                   &                  &                &     \\
                 &   L4 & 2       & 9.21/5.20        & 11.45/5.46      & 4.91/2.52     & 8.52/4.39 \\
                 &      & 3       &4.46/1.61         & 2.96/1.31       & 1.07/0.34     & 2.83/1.09 \\
                 &      & 4       &4.34/1.57         & 2.99/1.33       & 1.14/0.35     & 2.82/1.08 \\
                 &      & 5       &12.27/9.66        &11.25/6.20       & 5.89/3.87     & 9.80/6.58\\
                 &      & 6       &16.20/9.90        & 16.13/8.32      & 9.70/6.00     & 14.01/8.07\\
                 &      & 7       & 15.88/9.85       & 15.49/8.27      & 8.91/5.91     & 13.43/8.01 \\
                 &      & 8       & 11.60/9.07       & 10.13/5.88      &4.96/3.68      & 8.90/6.21 \\
                 &      & 10      & 5.98/3.61        & 7.44/3.91       & 2.52/1.35     & 5.31/2.96 \\
                 &      & 12      & 5.15/2.59        & 7.00/3.37       & 2.02/1.00     & 4.72/2.32 \\
                 &      & 15      & 4.44/2.39        & 5.89/2.93       & 1.91/0.79     & 4.08/2.04\\
                 &      & 20      & 4.00/2.00        & 5.52/2.71       & 1.57/0.68     & 3.70/1.80\\
                 &      & 40      & 4.28/1.98        & 4.87/2.39       & 1.51/0.65     & 3.55/1.67 \\  
                \hline  
\end{tabular}
\end{center}
\label{table:uTCL}
}
\end{subtable}
\label{table:uTCL_stcl}
\end{table*}

For BN feature extraction, DNNs are trained  using data from the RSR2015 \cite{RSR2015} database, \emph{from which the pass-phrases that also appear in the TD-SV evaluation set in the RedDots database are removed}. Therefore, there are no pass-phrase overlap between data for training BN-DNNs and data for TD-SV evaluation.  It gives $\approx 72764$ utterances  over $27$ pass-phrases  (recorded in $9$ sessions) from $300$ non-target speakers ($157$ male, $143$ female). All DNN consists of $7$ layer feed-forward networks and use the same learning rate and the same number of epochs in training. Each hidden layer consists of $1024$ sigmoid units. The input layer is of $627$ dimensions, based on $57$ dimensional MFCC features with a context window of $11$ frames (i.e. $5$ frames left, current frame, $5$ frames right).%\\

For speaker-discriminant DNN (SPK-BN), the number of output nodes is equal to the number of speakers, i.e. $300$.  Whereas, the speaker+pass-phrase (SPK+phrases-BN) discriminant DNN consists of $327$ output nodes ($300$ speakers +  $27$ pass-phrases). To obtain the final BN feature, the output from a hidden layer, a $1024$ dimensional deep feature, is projected onto a $57$ dimensional space to align with the dimension of the MFCC feature for a fair comparison. Allowing a higher dimension for BN can potentially boost the performance as observed in \cite{yu2017adversarial}. Deep features are normalized to zero mean and unit variance at utterance level before using principle component analysis (PCA) for dimension reduction.

A gender-independent GMM-UBM with $512$ Gaussian components having a diagonal covariance matrix is trained using the $6300$ utterances  from 630 non-target speakers ($438$ male, $192$ female)  of the TIMIT database \cite{Timit}. Same GMM-UBM training data are used for the PCA. In MAP adaptation, three iterations are followed with value of relevance factor $10$. 

For the i-vector method, the data for training BN-DNNs are also used for training a gender independent total variability space and for training PLDA and sph. In PLDA, utterances of the same pass-phrase from a particular speaker are treated as an individual speaker. It gives $8100$ classes (4239 male and 3861 female) in PLDA. Speaker and channel factors are kept full in PLDA, i.e. equal to the dimension of i-vector ($400$) for all systems.

System performance is evaluated in terms of equal error rate (EER) and minimum detection cost function (minDCF) \cite{DET97}.

\section{Results and Discussions}
This section presents the TD-SV results for different features, followed by discussions. 

\subsection{Comparison of TD-SV performance for a number of BN features and MFCCs under the GMM-UBM framework}
In this section, we present TD-SV results of sTCL and uTCL with or without clustering, using 10 TCL classes and extracting features from BN-DNN hidden layers L2 and L4, as well as TD-SV results of phone-discriminant BN features. We compare these results to those of speaker-discriminant BN features and MFCCs. 

Table \ref{table:table1} shows the TD-SV results of different BN features and MFCCs. 
It is noticed that all BN features (except for PHN-BN1-L4, sTCL-L4 and uTCL-L4, but L2 should be used for these methods as the training targets are phonetic content-related) give lower average EERs and MinDCF than those of MFCCs, confirming the effectiveness of BN features for the TD-SV. The behavior of sTCL-L4 and uTCL-L4 is analyzed and discussed in the next subsection. Concerning the hidden layer from which features are extracted, L$4$ is be
tter than L$2$ for SPK-BN and SPK+phrase-BN, while the opposite is observed for the rest, including sTCL-BN, uTCL-BN, PHN-BN1, PHN-BN2 and PHN-BN3. This can be well explained by the fact that the training target classes include speaker identities for SPK-BN and SPK+phrase-BN and thus using later hidden layer as output is favourable. 

Among all features without clustering, PHN-BN3 gives the lowest average EER followed by uTCL-BN. Among PHN-BN features, the ranking in TD-SV performance is PHN-BN3, PHN-BN2 and PHN-BN1, in a descending order. This is also in line with their speech recognition performance as PHN-BN3 uses the forced-alignment decoding approach and thus provides the most accurate phonetic transcriptions for training DNNs. 

The clustering method is able to reduce the the average EER and MinDCF of PHN-BN1 and PHN-BN2 with respect to their standalone systems. %This signifies that clustering helps by better grouping the similar data/events on the speech signal within the group where they are supposed to be in sense of maximum likelihood sense in unsupervised manner. 
However, it is unable to improve the performance of PHN-BN3. This is because the already accurate transcriptions provided by the forced-alignment decoding approach. 

Among all the feature extraction methods, uTCL-BN with clustering gives the lowest average EER and minDCF, followed by PHN-BN3 with a minor margin. 
% This further demonstrate that TCL+clustering is beneficial for the TD-SV and they are able to leverage the large amount of pass-phrase (without label information) available in the real-world for the machine learning.

\subsection{TD-SV performance of TCL-BN features with different configurations under the GMM-UBM framework}
Table \ref{table:uTCL_stcl} presents TD-SV results of sTCL and uTCL with or without clustering, using different numbers of TCL classes and extracting features from different BN-DNN hidden layers with the purpose of providing insights about the behaviour of TCL with different configurations. 

We first compare the performance of extracting features from different hidden layers for sTCL and uTCL. L2 clearly outperforms L4. This can be explained by the fact that the TCL training target classes are related more to phonetic content than to speaker identity, so that the earlier output layer is preferred for speaker verification. The differences between L2 and L4 for sTCL are marginal, while the differences for uTCL are very significant. The performances of sTCL do not change much across different numbers of training target classes and different layers (L2 or L4), and they are all better than the MFCC baseline. This stable performance of sTCL is primarily due to the fact that sTCL randomly assigns labels to segments. On the other hand, the performance of uTCL varies much. An overall explanation to these observations is that the training targets for uTCL are much more meaningful and consistent than those for sTCL. 

Concerning the number of TCL classes, $N=15$ and $N=10$, give the lowest average EERs for sTCL and uTCL, respectively. The performance of sTCL does not vary much for different numbers of classes, which is due to the nature of sTCL randomly generating segments and assigning class labels. On the other hand, uTCL is rather sensitive to varying the value $N$. Different from sTCL, uTCL exploits the data structure of text-dependent pass-phrases, which is the reason why it is sensitive to the number of classes.
% For $L2$, smaller number of classes is better while it is the opposite for $L4$, and the differences are significant.  When $N=5$ or $N=10$, it is highly likely that uTCL models the phonetic information well . 

\begin{figure}[t]
\hspace*{+2.0cm}\includegraphics[width=15.4cm,height=8.0cm]{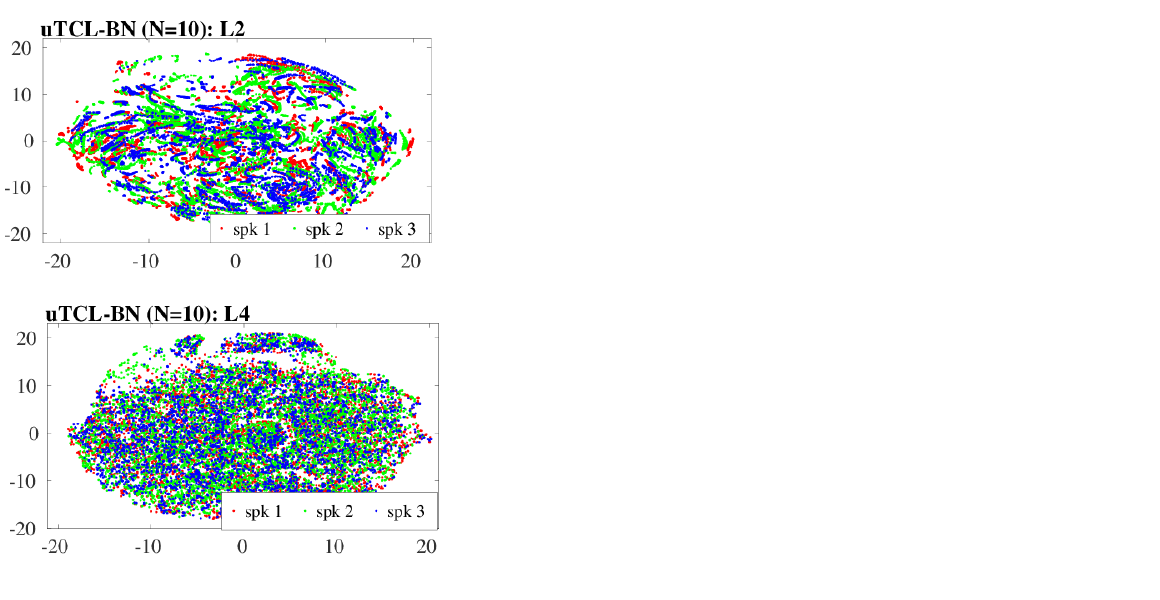} 
\caption{\it Scatter plots of uTCL-BN-features for the L2 and L4 DNN layers. The plots are extracted for three target speakers using the utterances available in the training set (using the T-SNE toolkit \cite{tsne} with same parameters). All features use the same utterances of the three speaker for a better comparison.}
\label{fig:scatter_plot_L2_L4_N10}
\end{figure}

The behaviour of uTCL deserves extra attention. When the number of classes $N$ equals to 10, uTCL-L2 achieves the lowest EER $(1.89\%)$ and MinDCF $(0.76/100)$ while uTCL-L4 gives the second highest EER $(16.61\%)$ and the third highest MinDCF $(8.59/100)$, among all configurations without clustering, and the differences are large. 
On the other hand, $N=7$ gives the worst performance (still slightly better than the MFCC baseline) among uTCL-L2 while the third best among uTCL-L4. The exactly opposite performance between L2 and L4 is an interesting observation.  To provide an insight about this behaviour, we scatter-plot the uTCL-BN features for the L2 and L4 layers for $N=10$ using the T-SNE toolkit \cite{tsne}, as shown in Fig.\ref{fig:scatter_plot_L2_L4_N10}. From the Fig. \ref{fig:scatter_plot_L2_L4_N10}, it can be seen that uTCL-L4 BN features all mixed together and does not show any discrimination structure or pattern in the feature space. On the other hand, uTCL-L2 features form clusters for different speakers. This reflects on their performance of TD-SV. 
% For uTCL-L2, N=10 gives the best performance (much better than the MFCC baseline) while N=7 the worst (similar to the MFCC baseline), while it is the opposite for uTCL-L4, which is an interesting observation. For example, when N=10, the \%EER results for uTCL-L2 and uTCL-L4 are 1.89 and 16.61, respectively. To further investigate why this big difference, 

Similar behaviour to that of $N=10$ is observed for $N=5$. This is likely because $N=5$ and $10$ match the underline linguistic structure of utterances in the RSR2015 database so that $L4$ strongly represents the linguistic information and the network learns good feature representation for speech signal in general at L2. Analysis shows that the minimal, maximal and average number of words per sentence in the database are 4, 8 and 6.3. Average number of frames per utterance is 205, and average number of frames per word is 32.5. Table \ref{table:uTCL_stcl} shows that $N=7$ and $N=15$ behave in an opposite way to that of $N=5$ and $N=10$, which deserves further investigation. 

% How about this observation for the higher value of N? \\
For larger values of $N$, e.g. $20$ and $40$, Table \ref{table:uTCL_stcl} shows that the differences in TD-SV performance among sTCL-L2, sTCL-L4, uTCL-L2 and uTCL-L4 are rather small, with EER ranging from $2.86\%$ to $3.00\%$, which are rather consistent but higher than that $(1.89\%)$ of uTCL-L2 for $N=10$. This is because small segments resulted from large N values increase the mismatch among segments with the same label. When $N=40$, the average  number of frames per segment is around 5, so it is more likely segments in the same class have different phonetic contents, leading to less-well trained BN-DNN as compared with smaller values of $N$, e.g. $N=10$, as well as leading to similar performances between sTCL and uTCL for L2 and L4. On the other hand, clustering helps improve the performance of uTCL-L2 much, by giving decent performances ($1.79\%$, $1.88\%$ and $2.15\%$ for $N=10$, $20$, and $40$ respectively). 
% the performances for different layers for both sTCL and uTCL are close to each other. When L4 is used, average error rates in TD-SV become much smaller, except for L4 without clustering. For L2 with clustering, the performances are decent. 
% specially for the use of higher layer of DNN in uTCL for the BN feature extraction. 
% It is expected that higher value of $N$ yields less number of frames per segment i.e. reduces overlap (confusion)  of the event across the speech signal. Besides, higher value of $N$ increases the number of classes in the output layer of DNN to be discriminated. As we are not  using any label information, so more number of classes consider the various kind of events/activities as withstand i.e. reduces the chance of mixing of different events across the data with respect to the system using lower value of $N$.

The clustering method steadily improves the performance of both sTCL and uTCL for L2. This indicates that the proposed clustering method is able to assign similar speech segments to the same class in an unsupervised manner. In other words, DNNs get better labelled data and thus reduce intra-class variabilities for DNN training, leading to better BN features for TD-SV. It is worth to note that after applying the clustering method, uTCL-L2 provides both stable and good performance across the different numbers of classes ranging from 5 to 20, which largely improves the applicability of uTCL.

% Less classes lead to smaller intra class variability. 
% As mentioned above, uTCL performs the best at $N=10$ for $L2$ and the worst for $L4$. When N=15 or 20, uTCL performs similarly to sTCL. 
% This could be because uTCL is capable of capturing the phonetic information through exploiting the text-dependent pass-phrase data in DNN training. It could be due to the reason that more number of classes in uTCL splits the data (supposed to be within a single group) into the others, which could reduce the intra class variabilities during the optimization of DNN. Whereas, data in the sTCL are already randomize and structure of the phonetic classes are messed up and hence performance of the TD-SV for the  increases number of TCL classes does not impact. 

It is observed that uTCL-L2 with clustering performs steadily well when the number of training target class is equal to or larger than the average number of words in utterances and it performs the best at around two times the average number of words. 

It should be noted that in all experiments in this work, the pass-phrases in the DNN training data are different from  the TD-SV evaluation set, i.e. the learned feature is not phrase-specific.

\begin{figure*}[t]
\hspace*{+0.1cm}
\includegraphics[width=11.4cm,height=7.0cm]{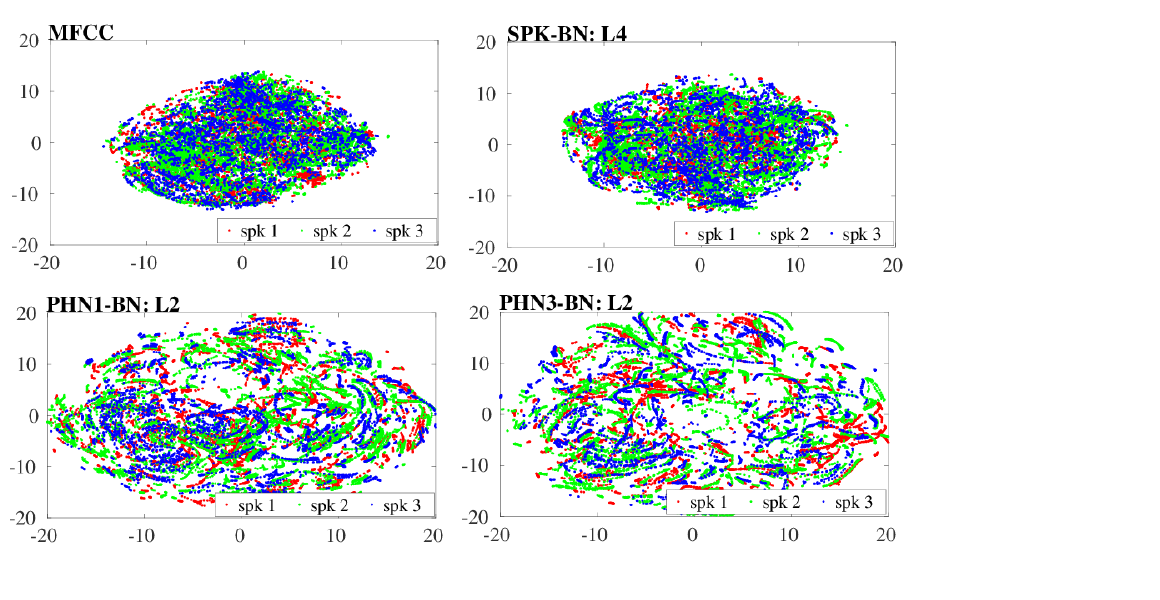} \label{fig:MFCC_spkr}
%\vspace*{-0.5cm}
\hspace*{-2.50cm}
\includegraphics[width=11.4cm,height=7.0cm]{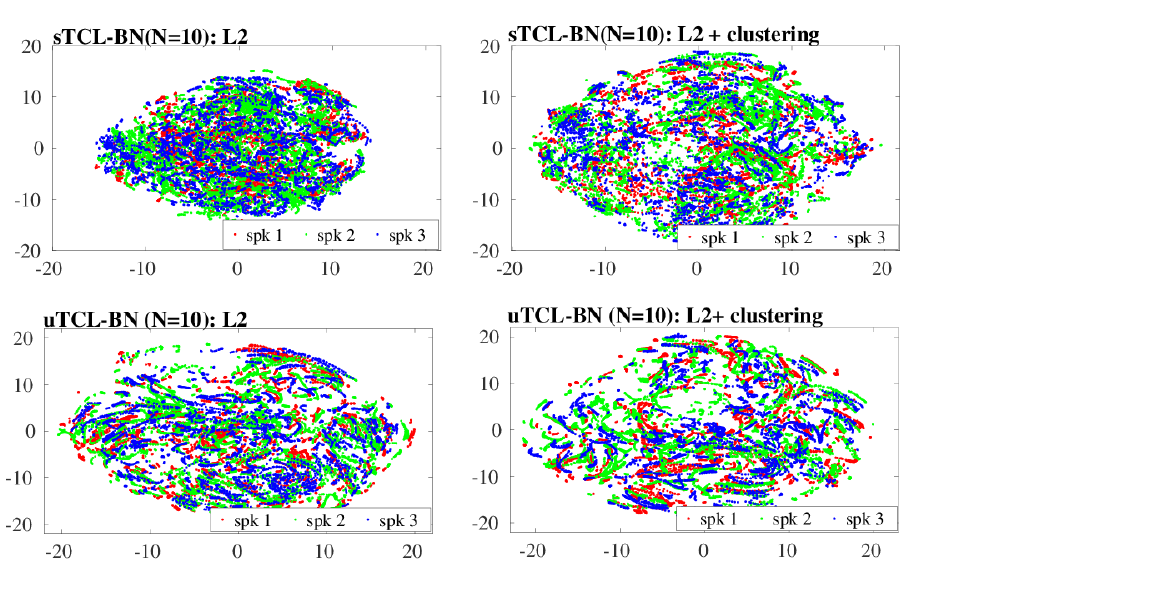} \label{fig:uTCL}
\caption{\it Scatter plots of MFCCs and BN-features extracted for three target speakers using the utterances available in the training set (using T-SNE toolkits \cite{tsne} with same parameters). All features use the same utterances of the three speaker for a better comparison.}
\label{fig:scatter_plo}
\end{figure*}

\begin{table*}[t]
%\scriptsize{
\caption{\it TD-SV results of MFCCs and BN features on the m-part-01 task of the RedDots database using the i-vector method}
\begin{center}
\begin{tabular}{|lcccccc|c|}\cline{1-8}
 Feature     & DNN    & \# of  & Clustering   &\multicolumn{3}{c}{Non-target type [\%EER/(minDCF$\times$ 100)]} & Average\\ 
             &Lyr.    & classes& without: $\times$   &Target-   & Impostor-        & Impostor-     & (EER \\
              &       &        & with: \checkmark  &wrong        & correct      & wrong          &  /minDCF)   \\  \hline \hline  
 MFCC         &  -    & -      &  - & 6.96/3.23    & 4.82/2.03    & 1.63/0.61      & 4.47/1.96 \\ 
              &       &        &    &           &              &                &     \\ 
 SPK-BN      &  L4   & 300     & -  & 7.19/3.02   & 5.76/2.29    & 2.33/0.81      & 5.10/2.04 \\
              &       &        &    &          &              &                &     \\
SPK+phrase-BN& L4  & 327       & -  & 7.27/3.01    & 6.07/2.34    & 2.11/0.85     & 5.15/2.02 \\
               &      &       &    &          &              &               &  \\
PHN-BN1        & L2   &  38 & $\times$    & 2.68/1.04    & 4.57/1.94    & 0.89/0.26     & 2.71/1.08 \\        
               &      &    &  \checkmark &  2.76/1.31    & 4.13/1.79     & 0.67/0.25 & {\bf 2.52}/1.12  \\          
               &     &     &             &   &              &                &     \\
PHN-BN2     & L2 & 47     & $\times$        & 2.87/1.15    & 4.71/1.86    & 0.89/0.30      & 2.83/1.10  \\ 
            &    &     & \checkmark & 2.37/1.03      & 3.93/1.79    & 0.89/0.25     & {\bf 2.40}/1.02 \\ 
            &        &    &     &              &              &                &     \\
PHN-BN3 (ASR force-alignment) & L2 & 39   & $\times$ & 2.25/0.83    & 4.65/1.90      & 0.89/0.26  & {\bf 2.59}/1.00   \\
            &    &      & \checkmark &  2.89/1.16    & 4.16/1.82      & 0.92/0.31  & 2.66/1.10  \\ 
                   &    &      &     &          &              &                &     \\
                   
sTCL-BN            &  L2  & 10 & $\times$ & 6.60/2.97    & 5.51/2.25    &  1.80/0.74     & 4.63/1.99 \\
%& L2 & 15   & 7.24/3.05    & 5.73/2.28    & 1.93/0.78      & 4.97/2.04 \\  
                &    &      & \checkmark & 3.92/1.67    & 4.31/1.82    & 1.07/0.38      &  3.10/1.29  \\
%&    &    & 4.22/1.63    & 4.78/1.86    & 1.05/0.40      & {\bf 3.35}/1.30 \\
                   &    &      &   &           &              &                &   \\
uTCL-BN            & L2 & 10   & $\times$ & 2.74/0.97    & 5.27/2.08    & 0.95/0.32   & 2.991.12  \\
  &     &     & \checkmark &  2.73/1.11     & 4.19/1.86  & 0.92/0.27   & {\bf 2.61}/1.08  \\ \hline

\end{tabular} 
\end{center}
\label{table:ivector}
%}
\end{table*}

\subsection{Scatter plots of BN features and MFCCs}
To obtain insights about the different features, we use T-SNE toolkits \cite{tsne} to scatter-plot the different features for $3$ target speakers (to limit the number for better visualization) using the utterances available in the training set as in Fig.\ref{fig:scatter_plo}. It can been seen that MFCC features are more compact and mixed together with each other. SPK-BN is slightly better, but not significantly. On the contrary, PHN-BN3 and uTCL+clustering BN features are much more spread and demonstrate clear structures in the data, indicating the superior discrimination and representation ability. It is further noticed that clustering helps make the TCL features more spread and structured. It is encouraging to see that the level of spread and structure of features is well in-line with their corresponding performance in TD-SV. This indicates that the scatter plot generated by using T-SNE is a good means for choosing features and thus the configurations to generate the features.   
% From the Fig.\ref{fig:scatter_plo}, it can be seen that data-points among the speakers are more discriminate (as well low intra variability within the same speakers) in the \emph{sTCL}, \emph{uTCL} and \emph{PHN} based systems with compared to the MFCCand \emph{SPK-BN}. And s/u TCL+clustering yields more discriminate features than the standalone sTCL, which reflects the significant gain in EER value of TD-SV. With respect to ASV based systems, uTCL with/without clustering shows very similar patterns on the data-points among the speakers and those system performances are very closer to each others.    

\subsection{Comparison of TD-SV performance for a number of BN features and MFCCs under the i-vector framework}
Table \ref{table:ivector} compares the TD-SV performance of several features under the i-vector framework \cite{Deka_ieee2011} on the m-part-01 task of the RedDots database. For simplicity, we only consider the DNN layer for BN feature extraction, which gives the lowest average EERs in Table \ref{table:table1}.  It can be seen from the Table \ref{table:ivector} that average EER or MinDCF values of the TD-SV for most of BN features are lower than those of MFCCs except for SPK-BN and SPK+phrase-BN. This again confirms the usefulness of BN features for TD-SV. Among all features, PHN-BN2 with clustering performs the best, followed by PHN-BN1 with clustering. PHN2-BN and uTCL-BN with clustering come after with small margins. It is interesting to notice that it is not the one with most accurate transcriptions gives the best TD-SV performance under the i-vector framework, even though the margins are small. Compared to the GMM-UBM framework with results shown in Table \ref{table:table1}, the i-vector method gives much higher EER and minDCF values. This is due to the use of short utterances for speaker verification \cite{Yuan2015,Delgado2016Asru,RSR2015}.

\begin{table*}
\caption{TD-SV results for the score-level fusion of MFCCs and BN features on the m-part-01 task of the RedDots database using the GMM-UBM method}
\begin{center}
%{\small
\begin{tabular}{|lccc|c||c|}\cline{1-6}
 Score fusion & \multicolumn{3}{c}{Non-target type [\%EER/(MinDCF$\times$ 100)]} & Average      & Without fusion\\
  $(\# no. of classes)$         & Target-wrong  & Impostor-correct & Impostor-wrong & EER/MinDCF & Avg.EER/MinDCF   \\ \hline \hline 
  MFCC                           & 5.12/2.17    & 3.33/1.40    & 1.14/0.47   & 3.19/1.35 &  3.19/1.35 \\
  MFCC \& SPK-BN($300$)              & 4.59/1.72  & 2.74/1.19    & 0.89/0.33  & {\bf2.74/1.08} & 2.91/1.13 \\
  MFCC \& SPK($300$)+phrase($27$)-BN & 4.62/1.70  &2.77/1.20     & 0.86/0.33  & {\bf2.74/1.08} &  2.92/1.12\\
  MFCC \& PHN-BN1 ($38$) + clustering  & 2.56/0.85  & 2.69/1.15   & 0.57/0.17 & {\bf1.94}/0.72  & 1.97/0.72\\
  MFCC \& PHN-BN2 ($47$) + clustering  & 2.34/0.86  & 2.43/1.13   & 0.61/0.21 & {\bf 1.80/0.73} & 1.81/0.74 \\
  MFCC \& PHN-BN3 ($39$) + clustering  & 2.25/0.79  & 2.49/1.11   & 0.56/0.16 & {\bf1.77}/0.69  &1.82/0.69 \\
  MFCC \& sTCL-BN($N=10$) + clustering & 3.14/1.21 & 2.56/1.20    & 0.77/0.25  & {\bf2.15}/0.89 & 2.23/0.87 \\
  MFCC \& uTCL-BN ($N=10$) + clustering &2.06/0.71 &2.54/1.10     & 0.59/0.17  & {\bf1.73}/0.66 & 1.79/0.65   \\ \hline 
\end{tabular}
%}
\end{center}
\label{table:eval_score_fusion}
\end{table*}

\subsection{Fusion of MFCCs with BN features}
In this section, we study the fusion of MFCCs and BN features at both score and feature levels under the GMM-UBM framework.
Only the GMM-UBM framework and BN features with clustering are considered due to their good performance. 

\subsubsection{Score-level fusion}
Table \ref{table:eval_score_fusion} presents the TD-SV results when scores of the MFCC based system are fused with the scores of the respective BN feature based systems.  
% As it is observed in earlier tables \ref{table:table1}-\ref{table:ivector} that GMM-UBM technique and BN feature with clustering  yield the better SV performance than the i-vector and system without clustering, respectively.  
Scores of the different systems are combined with weights as follows. First, the inverse of the mean EER value ($m_{eer}^i$) of each system $i$ is calculated. Second, inverse values are scaled so that the summation of the weights ($w_i$ for the $i^{th}$ system) become unity. Finally the fusion score is the weighted sum of component system scores. The steps are detailed in the following equations.   
\begin{eqnarray}
y_i &=&  \frac{1}{m_{eer}^{i}} \\
w_i &= & \frac{y_i}{\sum_{i=1}^{l} y_i} \\
fused_{score} &=& \sum_{i=1}^{l} w_i*score_{sys_{i}}
\end{eqnarray}
From Table \ref{table:eval_score_fusion}, it is noticed that all fusion systems perform better than MFCCs alone. When combined with MFCCs, all BN features obtain better performance compared to their standalone counterparts. This  shows that BN features carry  information complementary to MFCCs when used for TD-SV. uTCL-BN with clustering still gives the best performance followed by PHN-BN3. 
% However, TCL does not use any manual annotation/transcription (in PHN, ASR training needs annotated data). Therefore, TCL would be useful for the real-life for utilizing the huge amount of available pass-phrase based data for the TD-SV. Better fusion technique could further improve the performance of the TD-SV in respective systems. 

\subsubsection{Feature-level fusion}
\begin{figure}[h]
\includegraphics[width=8.8cm,height=6.8cm]{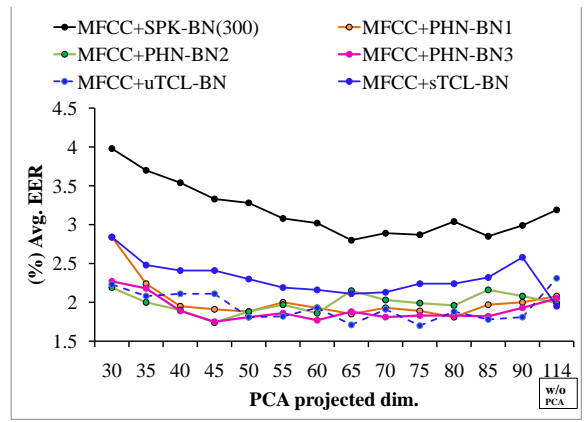} 
\caption{\it The TD-SV performance for various dimensions of PCA projected augmented feature (MFCC+BN) of different systems on the m-part-01 task of the RedDots database using GMM-UBM.}
\label{fig:PCA_eer}
\end{figure}

Fig.\ref{fig:PCA_eer} shows the TD-SV performance (average EER over target-wrong, imposter-correct and imposter-wrong cases) for  various dimension of PCA projected augmented feature (MFCC+BN) of different systems on the m-part-01 task of the RedDots database using the GMM-UBM. It is shown in \cite{Cong-Thanh2013} that simply augmenting features may degrade the performance due to the redundancy between the features. PCA is implemented as per \cite{Cong-Thanh2013}.  %\textcolor{red}{ \st{ where  PCA matrix is estimated by pooling the mean feature vector per speech file over many utterances. As deep features are normalized to zero mean and unit variance before projecting onto the PCA space (PCA also developed the using normalized deep features using TIMIT) to get BN. Therefore, mean value of all frames/feature vectors per speech file is close to zero.  So, we randomly down-sample the number of frames per  utterance and then a mean feature vector  is calculated on the selected frames for the particular  utterance (so that mean value of feature vectors for a utterance does not become zero value). Afterwards PCA matrix is estimated using the mean feature vectors of the utterances.}}
From Fig. \ref{fig:PCA_eer}, it can be observed that augmented feature +PCA gives slight reduction of average EER except for the SPK-BN(300) with respect to the system without PCA.  
%MFCCs and BN features are simply concatenated together, which gives $114$-dimensional features for TD-SV.
%Same as done in the score-level fusion, we present the results for BN features with clustering under the GMM-UBM framework, as detailed in \ref{table:augment_fusion}. 

\section{Conclusions}
In this paper, we presented a time-contrastive learning (TCL) based bottleneck (BN) feature extraction method for the text-dependent speaker verification (TD-SV). Specifically, a speech utterance/signal is uniformly partitioned into a number of segments of multiple frames (each corresponding to a class) without using any label information and then a deep neural network (DNN) is trained to discriminate speech frames among the classes to exploit the temporal structure in the speech signal. In addition, we proposed a segment-based clustering method that iteratively regroups speech segments to maximize the likelihood of all speech segments. It was experimentally shown that the proposed TCL-BN feature with clustering gives better TD-SV performance than Mel-frequency cepstral coefficients (MFCCs) and existing BN feature extracted by discriminating speakers or speakers and pass-phrases and it is further better than or on par with phone-discriminant BN (PHN-BN) features that we investigated in this work. The clustering method is able to improve the TD-SV performance for both TCL-BN and PHN-BN, except for the type of PHN-BN that relies on forced-alignment to generate transcriptions. All BN features are shown to be complementary to MFCCs when score-level fusion is applied. Overall, the work has shown the effectiveness of TCL approach for feature learning in the context of TD-SV and the usefulness of PHN-BN. 
Future work includes the investigation of using TCL for text-independent speaker verification. 
% Addition, we studied the performance of the TD-SV for augmentation of cepstral feature with BN  in both score and feature domain. It demonstrates the effectiveness of the proposed TCL-BN in TD-SV with respect to the existing BN features.  System performance are represented on the Reddots challenge 2016 database for TD-SV using short utterances. 

\bibliographystyle{IEEEbib}
\bibliography{strings,References}
\begin{IEEEbiography}[{\includegraphics[width=1in,height=1.25in,clip,keepaspectratio]{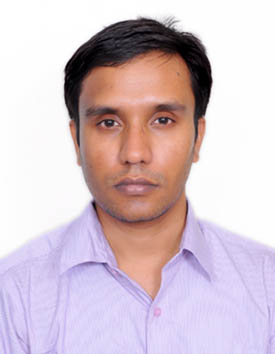}}]
{Achintya Kr. Sarkar} received the M. Tech. degree in Instrumentation and Control Engineering from Punjab Technical University, India, in 2006, and a Ph.D. degree from IIT Madras, Chennai, India, in 2011. From 2011 to 2017, he was working as a research fellow in several laboratories: Laboratoire Informatique d'Avignon, France,   LIMSI-CNRS, France,  Department of Electrical and Electronic Engineering, University College Cork, Ireland, and the Department of Electronic Systems, Aalborg University, Denmark. He is currently working as a faculty in the School of Electronics Engineering,  VIT-AP University, India. His research interests include  speech signal processing, biomedical signal processing, speaker recognition,  spoofing countermeasure, seizure detection and application of machine learning in the above areas.
\end{IEEEbiography}

\begin{IEEEbiography}[{\includegraphics[width=1in,height=1.25in,clip,keepaspectratio]{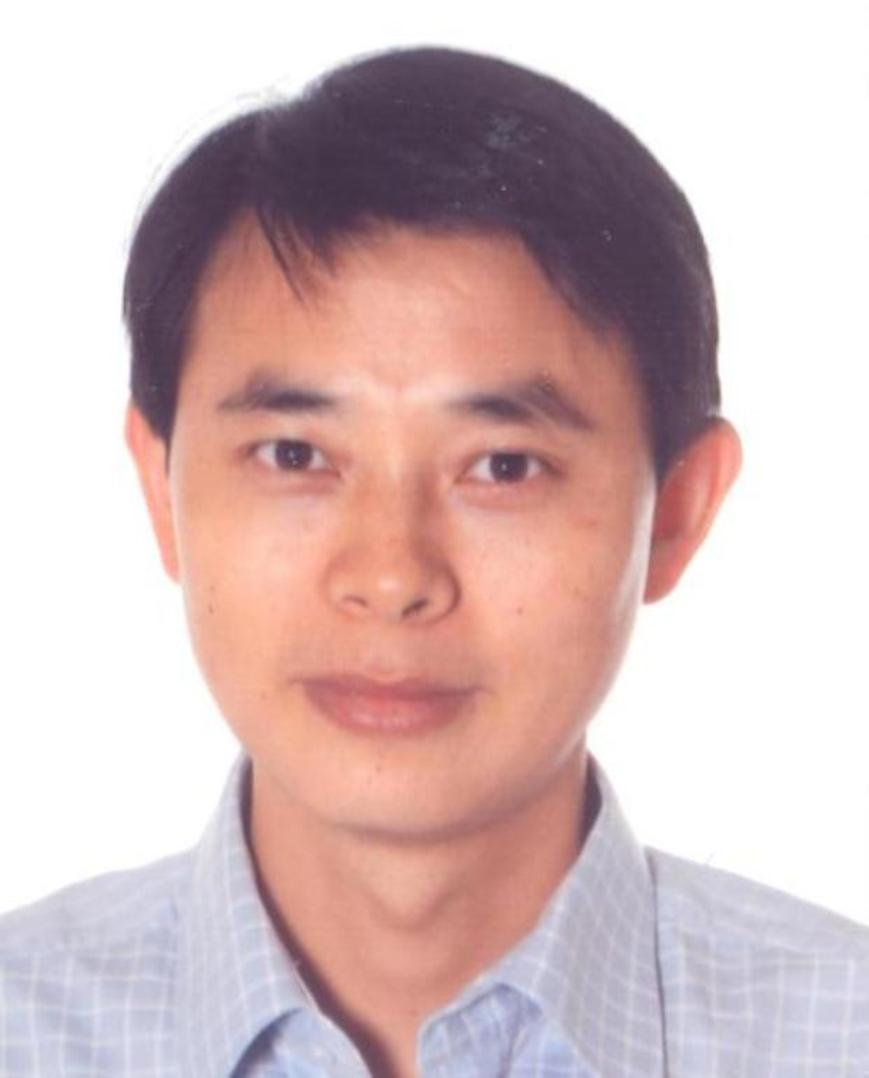}}]{Zheng-Hua Tan (M'00--SM'06)} received the B.Sc. and M.Sc. degrees in electrical engineering from Hunan University, Changsha, China, in 1990 and 1996, respectively, and the Ph.D. degree in electronic engineering from Shanghai Jiao Tong University, Shanghai (SJTU), China, in 1999. 

He is a Professor in the Department of Electronic Systems and a Co-Head of the Centre for Acoustic Signal Processing Research (CASPR) at Aalborg University, Aalborg, Denmark. He was a Visiting Scientist at the Computer Science and Artificial Intelligence Laboratory (CSAIL), Massachusetts Institute of Technology (MIT), Cambridge, USA, an Associate Professor in the Department of Electronic Engineering at SJTU, Shanghai, China, and a postdoctoral fellow in the Department of Computer Science at KAIST, Daejeon, Korea. His research interests include machine learning, deep learning, pattern recognition, speech and speaker recognition, noise-robust speech processing, multimodal signal processing, and social robotics. He has authored/coauthored about 200 publications in refereed journals and conference proceedings. He is a member of the IEEE Signal Processing Society Machine Learning for Signal Processing Technical Committee (MLSP TC). He has served as an Editorial Board Member/Associate Editor for Computer Speech and Language, Digital Signal Processing, and Computers and Electrical Engineering. He was a Lead Guest Editor of the IEEE Journal of Selected Topics in Signal Processing and a Guest Editor of several journals including Neurocomputing. He is the General Chair for IEEE MLSP 2018 and was a Technical Program Co-Chair for IEEE Workshop on Spoken Language Technology (SLT 2016).
\end{IEEEbiography}

 %\vfill
\begin{IEEEbiography}[{\includegraphics[width=1in,height=1.25in,clip,keepaspectratio]{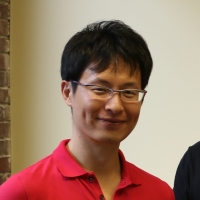}}]
{Hao Tang} is a postdoctoral associate at Massachusetts Institute of Technology. He obtained his Ph.D. from Toyota Technological Institute at Chicago in 2017 and his M.S. and B.S. from National Taiwan University. His research focuses on machine learning and its application to speech processing. His recent work includes segmental models, domain adaptation, end-to-end training, and speech representation learning.
\end{IEEEbiography}
\begin{IEEEbiography}[{\includegraphics[width=1in,height=1.25in,clip,keepaspectratio]{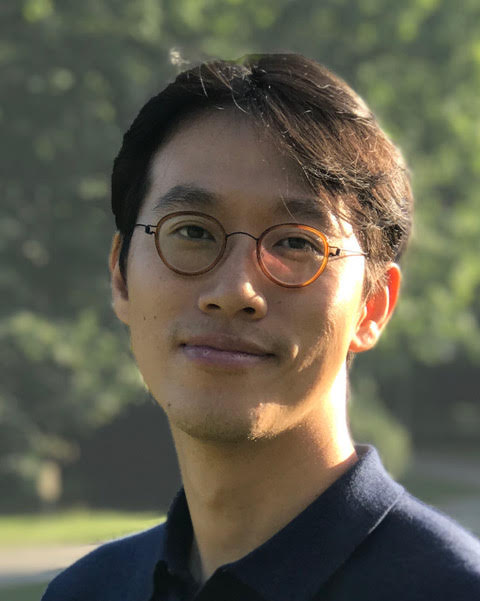}}]
{Suwon Shon}received B.S and Integrated Ph. D degree on electrical engineering from Korea University, South Korea in 2010 and 2017, respectively. From 2017, he joined Massachusetts Institute of Technology, MA, USA and working as post-doctoral associate at Computer Science and Artificial Intelligence Laboratory. His current research interests include the areas of automatic speech/speaker recognition, language/dialect identification and multi-modal recognition.
\end{IEEEbiography}

\begin{IEEEbiography}[{\includegraphics[width=1in,height=1.25in,clip,keepaspectratio]{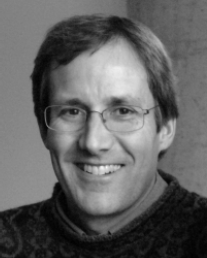}}]
{James Glass (F'14)} is a Senior Research Scientist at MIT where he leads the
Spoken Language Systems Group in the Computer Science and Artificial
Intelligence Laboratory.  He is also a member of the Harvard-MIT
Health Sciences and Technology Faculty.  Since obtaining his S.M. and
Ph.D. degrees at MIT in Electrical Engineering and Computer Science,
his research has focused on automatic speech recognition, unsupervised
speech processing, and spoken language understanding.  He is an IEEE
Fellow, and a Fellow of the International Speech Communication
Association, and is currently an Associate Editor for Computer,
Speech, and Language, and the IEEE Transactions on Pattern Analysis 
and Machine Intelligence.
\end{IEEEbiography}
\end{document}